\documentclass[preprint,12pt,authoryear]{elsarticle}

\usepackage{amssymb}
\usepackage{xcolor}
\usepackage{tablefootnote}
\usepackage{amsmath}

\usepackage{lineno}

\journal{Icarus}

\newcommand{\altwosix}{\mbox{${}^{26}{\rm Al}$}}
\newcommand{\mnfivethree}{\mbox{${}^{53}{\rm Mn}$}}
\newcommand{\hfoneeighttwo}{\mbox{${}^{182}{\rm Hf}$}}
\newcommand{\fesixty}{\mbox{${}^{60}{\rm Fe}$}}
\newcommand{\alratio}{\mbox{${}^{26}{\rm Al}/{}^{27}{\rm Al}$}}
\newcommand{\feratio}{\mbox{${}^{60}{\rm Fe}/{}^{56}{\rm Fe}$}}
\newcommand{\beratio}{\mbox{${}^{10}{\rm Be}/{}^{9}{\rm Be}$}}
\newcommand{\mnratio}{\mbox{${}^{53}{\rm Mn}/{}^{55}{\rm Mn}$}}
\newcommand{\hfratio}{\mbox{${}^{182}{\rm Hf}/{}^{180}{\rm Hf}$}}
\newcommand{\pdratio}{\mbox{${}^{107}{\rm Pd}/{}^{108}{\rm Pd}$}}
\newcommand{\uratio}{\mbox{${}^{238}{\rm U}/{}^{235}{\rm U}$}} 
\newcommand{\chisq}{\mbox{$\chi_{\nu}^{2}$}}
\newcommand{\permil}{\mbox{\text{\textperthousand}}}
\newenvironment{Steve}{\color{black}}{\ignorespacesafterend}

\newenvironment{DRD}{\color{black}}{\ignorespacesafterend}
\newenvironment{EDIT}{\color{black}}{\ignorespacesafterend}

\graphicspath{{./}{figures/}}

\begin{document}

\begin{frontmatter}



\title{Statistical Chronometry of Meteorites: II. Initial Abundances and Homogeneity of Short-lived Radionuclides}

\author[inst1]{Steven J. Desch}

\affiliation[inst1]{organization={School of Earth and Space Exploration, Arizona State University},
            addressline={PO Box 871404}, 
            city={Tempe},
            postcode={85287-1404}, 
            state={Arizona},
            country={USA}}

\author[inst2]{Daniel R. Dunlap}
\affiliation[inst2]{organization={Oak Ridge National Laboratory}, 
            addressline={1 Bethel Valley Rd}, 
            city={Oak Ridge},
            postcode={37830}, 
            state={Tennessee},
            country={USA}}

\author[inst3]{Curtis D. Williams}

\affiliation[inst3]{organization={Earth and Planetary Sciences Department, University of California, Davis},
            addressline={One Shields Ave.}, 
            city={Davis},
            postcode={95616}, 
            state={California},
            country={USA}}
     
\author[inst4]{Prajkta Mane}

\affiliation[inst4]{organization={Lunar and Planetary Institute, USRA},
            addressline={3600 Bay Area Blvd.},
            city={Houston},
            postcode={77058},
            state={Texas},
            country={USA}}

\author[inst6]{Emilie T. Dunham}

\affiliation[inst6]{organization={Department of Earth, Planetary and Space Sciences, University of California, Los Angeles},
            addressline={PO Box 951567},
            city={Los Angeles},
            postcode={90095-1567},
            state={California},
            country={USA}}


\begin{abstract}
Astrophysical models of planet formation require accurate radiometric dating of meteoritic components by short-lived (Al-Mg, Mn-Cr, Hf-W) and long-lived (Pb-Pb) chronometers, to develop a timeline of such events in the solar nebula as formation of Ca-rich, Al-rich Inclusions (CAIs), chondrules, planetesimals, etc.
CAIs formed mostly around a time (``$t\!\!=\!\!0$") when the short-lived radionuclide $\altwosix$ ($t_{1/2} = 0.72$ Myr) was present and presumably homogeneously distributed at a known level we define as $(\alratio)_{\rm SS} \equiv 5.23 \times 10^{-5}$.
The time of formation after $t\!\!=\!\!0$ of another object can be found by determining its initial $(\alratio)_0$ ratio and comparing it to $(\alratio)_{\rm SS}$.
Dating of meteoritic objects using the Mn-Cr or Hf-W systems is hindered because the abundances $(\mnratio)_{\rm SS}$ and $(\hfratio)_{\rm SS}$ at $t\!\!=\!\!0$ are not known precisely.
To constrain these quantities, we compile literature Al-Mg, Mn-Cr, Hf-W and Pb-Pb data for 14 achondrites and use novel statistical techniques to minimize the discrepancies between their times of formation across these systems.
{\Steve We find that for 
$(\mnratio)_{\rm SS} = (8.09 \pm 0.65) \times 10^{-6}$, 
$(\hfratio)_{\rm SS} = (10.42 \pm 0.25) \times 10^{-5}$, 
$t_{\rm SS} = 4568.36 \pm 0.20 \, {\rm Myr}$, and
a $\mnfivethree$ half-life of $3.80 \pm 0.23$ Myr, these four free parameters make concordant 37 {\EDIT out of 38} formation times recorded by the different systems in 14 achondrites.}
These parameters also make concordant the ages derived for chondrules from CB/CH achondrites, formed simultaneously in an impact, {\Steve and are apparently concordant with the I-Xe chronometer as well}.
Our findings provide very strong support for homogeneity of $\altwosix$, $\mnfivethree$, and $\hfoneeighttwo$ in the solar nebula, and our approach offers a framework for more precise chronometry. 
\end{abstract}


\begin{highlights}
\item We present a new method for combining and averaging data from the Al-Mg, Mn-Cr, Hf-W, and Pb-Pb radiometric dating systems, to: attain greater accuracy and precision in the initial $(\mnratio)_{\rm SS}$ and $(\hfratio)_{\rm SS}$ ratios the Pb-Pb age $t_{\rm SS}$ of ``$t\!\!=\!\!0$" in the Solar System, when $(\alratio)_{\rm SS} \equiv 5.23 \times 10^{-5}$; and better assess concordancy.

\item In meteorites and components where it is expected, we find substantial concordancy between the times of formation measured by the different isotopic systems, provided {\Steve $t_{\rm SS} = 4568.36 \pm 0.20$ Myr, $(\mnratio)_{\rm SS} = (8.09 \pm 0.65) \times 10^{-6}$, and $(\hfratio)_{\rm SS} = (10.42 \pm 0.25) \times 10^{-5}$, and the $\mnfivethree$ half-life is $\approx 3.80 \pm 0.23$ Myr}; this strongly implies homogeneity of $\altwosix$, $\mnfivethree$, and $\hfoneeighttwo$ in the solar nebula from early times.
\end{highlights}

\begin{keyword}
Solar System formation 1530 \sep Planet formation 1241 \sep Meteorites 1038 \sep Achondrites 15 \sep Chondrites 228
\end{keyword}

\end{frontmatter}



\section{Introduction} \label{sec:intro}

\subsection{Times of Formation from the Al-Mg and Pb-Pb Systems}

To learn about the birth of planets and the events of the Solar System's first few million years, we study meteorites that bear witness to this era. 
It is especially important to constrain the times at which the components in meteorites formed, the times at which their parent bodies accreted and melted, and when these bodies collided.
Most importantly, it is vital to constrain the relative order or sequence of events within the solar nebula. 
The goal is to find the time $\Delta t$ after $t\!\!=\!\!0$ that an event occurred, where $t\!\!=\!\!0$ is a defined event or time in the Solar System history.

To obtain these times $\Delta t$, radiometric dating systems such as the Al-Mg system are employed. 
Ca-rich, Al-rich Inclusions (CAIs) are thought to be the first solids formed in the Solar System.
When they formed, they incorporated live $\altwosix$, a short-lived radionuclide (SLR) that decays to ${}^{26}{\rm Mg}$ with a half-life of 0.717 Myr, or mean-life $\tau_{26} = 1.034$ Myr \citep{AuerEtal2009,Kondev2021}.
Although extinct now, its one-time existence can be found by taking a linear regression of the measured values of $y = {}^{26}{\rm Mg}/{}^{24}{\rm Mg}$ and $x = {}^{27}{\rm Al}/{}^{24}{\rm Mg}$ isotopic ratios in different minerals within the same CAI.
The slope of this correlation, if it is linear, yields $(\alratio)_0$ within the CAI at the time it formed and achieved isotopic closure. 
A large fraction of CAIs appear to have formed from a reservoir with $\alratio$ near a canonical ratio that we take as $(\alratio)_{\rm SS} \equiv 5.23 \times 10^{-5}$ \citep{JacobsenEtal2008}.
This strongly suggests that $\altwosix$ was homogeneously distributed in the solar nebula from a very early time. 
Assuming homogeneity of $\altwosix$, the time of formation of a CAI, as recorded by the Al-Mg system, can be calculated as 
\begin{equation}
\Delta t_{26} = \tau_{26} \, \ln \left[ \frac{ (\alratio)_{\rm SS} }{ (\alratio)_0} \right].
\end{equation}
This provides a date of formation, relative to $t\!\!=\!\!0$, which for our purposes is defined to be that time in the solar nebula when $(\alratio) = (\alratio)_{\rm SS}$. 

An alternative method is to use the Pb-Pb system to calculate the absolute age of a sample. 
Measurement of the isotopic ratios $x = {}^{204}{\rm Pb}/{}^{206}{\rm Pb}$ and $y = {}^{207}{\rm Pb}/{}^{206}{\rm Pb}$ in different leachates, washes, or residues derived from acid dissolution of a sample can be linearly regressed; the intercept of this regression, combined with a measurement of  ${}^{238}{\rm U}/{}^{235}{\rm U}$ in the bulk sample, yields a number that is a function only of the age of the sample, which we denote $t_{\rm Pb}$.
As we discuss in a companion paper (\citealt{DeschEtal2023a}; hereafter Paper I), this absolute age by itself is not a quantity that astrophysical models of planet formation can make use of; what matters is the sequence of events in the first few Myr of the solar nebula, not how long ago that sequence took place. 
Moreover, due to uncertainties in the half-lives of ${}^{235}{\rm U}$ and ${}^{238}{\rm U}$, absolute ages are intrinsically uncertain by $\pm 9  (2\sigma)$ Myr \citep{TissotEtal2017}.
However, these systematic uncertainties largely cancel when taking the difference between two Pb-Pb ages, and typically the Pb-Pb system can be used as a {\it relative} chronometer with precision of 0.3-0.5 Myr determined solely by measurement uncertainties \citep{Amelin2006,TissotEtal2017}.
The Pb-Pb ages of samples can be converted into $\Delta t_{\rm Pb}$, the time of formation after $t\!\!=\!\!0$: 
\begin{equation}
\Delta t_{\rm Pb} = t_{\rm SS} - t_{\rm Pb}.
\end{equation}
Here, $t_{\rm SS}$ is the Pb-Pb age of a sample that would be found if it achieved isotopic closure at $t\!\!=\!\!0$ [when $\alratio$ = $(\alratio)_{\rm SS}$], assuming the same uranium half-lives that are typically assumed (703.81 Myr for ${}^{235}{\rm U}$ and 4468.3 Myr for ${}^{238}{\rm U}$; \citealt{JaffeyEtal1971}, \citealt{VillaEtal2016}).

The most commonly accepted way to determine the value $t_{\rm SS}$ is by direct measurement of the Pb-Pb ages of CAIs, which are presumed to have achieved isotopic closure of the Pb-Pb system at the same time ($t\!\!=\!\!0$) as the Al-Mg system. 
The most commonly cited value is that of \citet{ConnellyEtal2012}, who averaged data from four CAIs to find $4567.30 \pm 0.16$ Myr.
Some CAIs appear older.
\citet{BouvierWadhwa2010} found one CAI to have a Pb-Pb age of $4568.2 \pm 0.2$ Myr, although this was not based on a direct measurement of $\uratio$ in the sample.
\cite{BouvierEtal2011a} reported one with an age of $4568.0 \pm 0.3$ Myr, but not in the refereed literature. 
These suggest that perhaps not all CAIs achieved isotopic closure of the Pb-Pb system at the same time; perhaps none of them achieved isotopic closure at $t\!\!=\!\!0$.

One goal of Paper I was to determine $t_{\rm SS}$, not by appealing to direct measurements of $t_{\rm Pb}$ in CAIs, but through a statistical approach, finding the value of $t_{\rm SS}$ that minimized the differences between $\Delta t_{26}$ and $\Delta t_{\rm Pb}$ across a basket of appropriate samples
{\Steve that rapidly cooled and were not later disturbed.
If the $\Delta t_{26}$ and $\Delta t_{\rm Pb}$ formation times of such samples cannot be reconciled, then this falsifies the assumption underlying the use of Al-Mg systematics for chronometry, that $\altwosix$ was homogeneous.
If a range of values for $t_{\rm SS}$ does make the formation times concordant, this strongly supports SLR homogeneity.}
{\Steve Using Al-Mg formation times and Pb-Pb ages for seven rapidly cooled achondrites (D'Orbigny, SAH 99555, NWA 1670, Asuka 881394, NWA 7325, NWA 2976 and NWA 6704). 
 \citet{DeschEtal2023a} found that a range of values  $t_{\rm SS} = 4568.42 \pm 0.24$ Myr made the Al-Mg and Pb-Pb ages concordant, with $\Delta t_{26}$ and $\Delta t_{\rm Pb}$ agreeing within errors for each achondrite, and the fit was good in a statistical sense
($\chi_{\nu}^2 = 0.98$). 
Even though chondrules may be reset by transient heating events, their Pb-Pb and Al-Mg formation times are consistent within measurement errors using the same value of $t_{\rm SS}$.
The goodness-of-fit parameter was still a statistically significant $\chisq = 1.36$ (12\% probability).
These results could have falsified the hypothesis of homogeneous $\altwosix$ but did not. 
These findings strongly support homogeneity of $\altwosix$ and the concordancy of Al-Mg and Pb-Pb formation times.}

{\Steve
The goal of this paper is to determine whether we can extend these results to other isotopic systems, assessing whether other SLRs were homogeneously distributed, and whether the formation times derived from them are concordant with the Al-Mg and Pb-Pb formation times.
}

\subsection{Times of Formation from other Isotopic Systems}

\begin{center}
\begin{figure}[ht!]
\includegraphics[width=0.8\textwidth,angle=0]{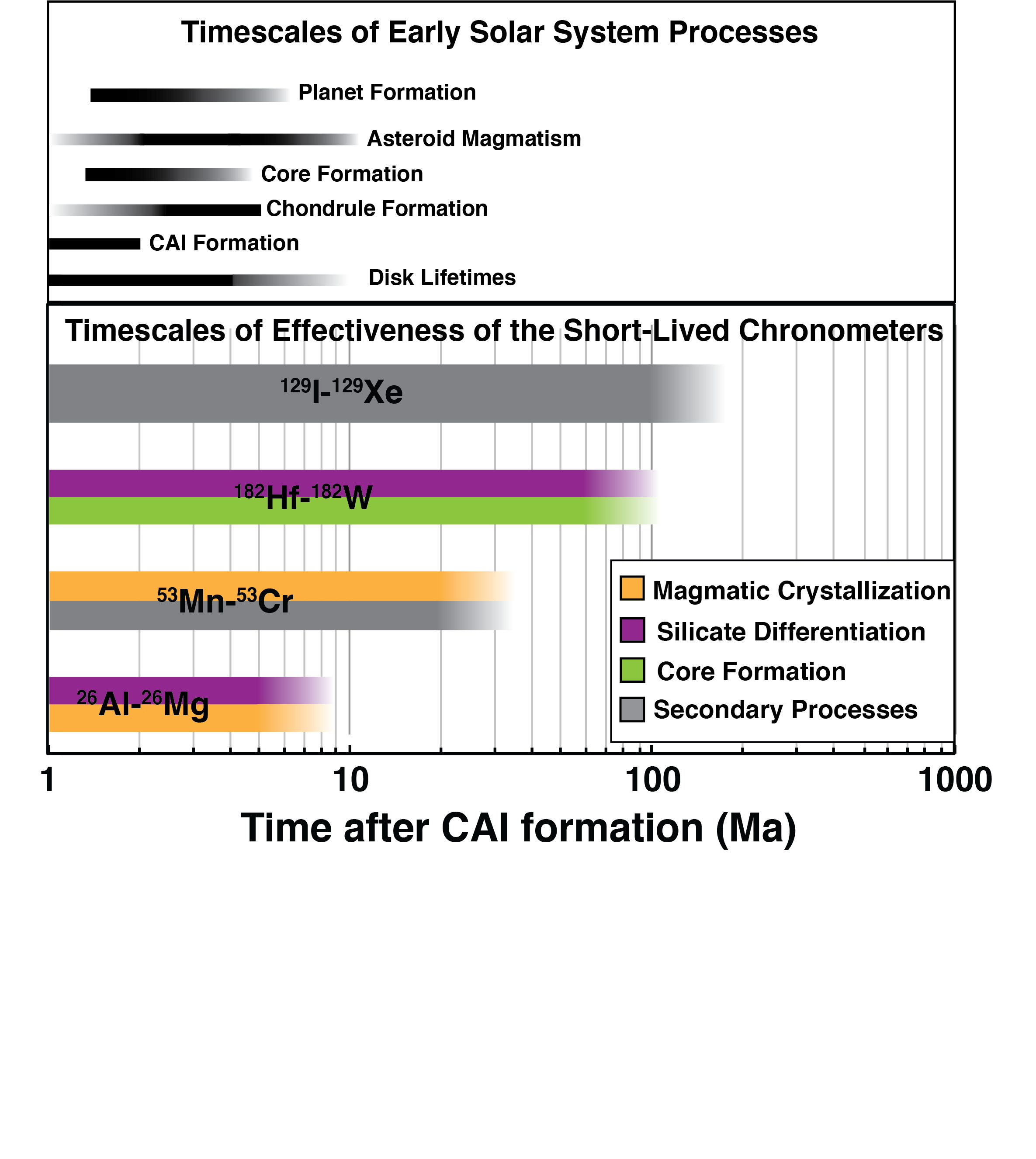}\vspace{-1.1in}
\caption{Timescales of effectiveness (roughly $8 \times$ half-life) of various short-lived chronometers, used to date early Solar System processes. The processes that affect isotopic closure and which are dated by each system are denoted by colors. The secondary processes relevant to ${}^{53}{\rm Mn}-{}^{53}{\rm Cr}$ and ${}^{129}{\rm I}-{}^{129}{\rm Xe}$ systems include aqueous and thermal alteration. The timescales of pertinent early Solar System processes are shown in the top panel.
\label{fig:chronometers}}
\end{figure}
\end{center}

{\Steve There are several other isotopic systems that can be used to date meteoritic samples, as depicted in {\bf Figure~\ref{fig:chronometers}}.
In general they date different sorts of events. 
The Al-Mg system usually achieves isotopic closure, at which point ${}^{26}{\rm Mg}$ ceases to diffuse significantly, after crystallization of rocky material from a magmatic melt.
Bulk excesses of ${}^{26}{\rm Mg}$ in a sample also can be used to date the time at which the sample became a closed reservoir, which often dates the time of silicate differentiation.
In practice, the Al-Mg system has been used to date these events as late as about 6 Myr, about 8 times the $\altwosix$ half-life.
}

The SLR $\mnfivethree$ decays to ${}^{53}{\rm Cr}$ with a half-life of about 3.7 Myr. 
The inferred initial ratio $(\mnratio)_0$ can be used to date the time of magmatic crystallization, or other processes such as carbonate formation. 
The SLR $\hfoneeighttwo$ decays (via ${}^{182}{\rm Ta}$) to ${}^{182}{\rm W}$ with a half-life of 8.9 Myr.
The initial $(\hfratio)_0$ ratio in a sample can be used to date magmatic crystallization or silicate differentiation, and excesses of ${}^{182}{\rm W}$ in bulk samples can be used to date metal-silicate separation, such as core formation.
The SLR ${}^{129}{\rm I}$ decays to ${}^{129}{\rm Xe}$ with a half-life of 16.1 Myr.
The initial $({}^{129}{\rm I}/{}^{127}{\rm I})_0$ ratio can be inferred and then used to date secondary processes such as shocks, as Xe tends to remain in a sample except when disturbed. 
Other SLRs that might be used as chronometers include ${}^{107}{\rm Pd}$, which decays to ${}^{107}{\rm Ag}$ with a half-life of 6.5 Myr; and ${}^{92}{\rm Nb}$, which decays to ${}^{92}{\rm Zr}$ with a half-life of 34.7 Myr.
Figure~\ref{fig:chronometers} depicts the timescales over which {\Steve some of} these chronometers are useful, as well as the types of processes that can be dated using them.
{\EDIT For more information, we refer the reader to the review by \citet{Davis2022}.}

In principle, a determination of an initial abundance when an object formed, such as $(\mnratio)_0$, could be used to determine the time of formation after $t\!\!=\!\!0$, which again we define as the time when $(\alratio) = (\alratio)_{\rm SS} \equiv 5.23 \times 10^{-5}$ in the solar nebula.
This time of formation as determined by Mn-Cr systematics would be 
\begin{equation}
\Delta t_{53} = \tau_{53} \, \ln \left[ \frac{ (\mnratio)_{\rm SS} }{ (\mnratio)_{0} } \right],
\label{eq:timeone}
\end{equation}
where $(\mnratio)_{\rm SS}$ would be the isotopic ratio in the solar system at $t\!\!=\!\!0$. 
Unlike the case for Al-Mg, for which a good estimate of $(\alratio)_{\rm SS}$ is known, the initial abundances like $(\mnratio)_{\rm SS}$ are for practical purposes not known to sufficient precision.
Direct determinations of $(\hfratio)_{\rm SS}$ and especially $(\mnratio)_{\rm SS}$ from CAI data yield values that are too uncertain to resolve times of formation at the $< 1$ Myr level; other isotopic ratios have hardly been constrained at all.  

As a result, meteoriticists have instead used these isotopic systems to measure the difference in times of formation between one object and a {\Steve {\bf single}} separate ``anchor" such as the achondrite D'Orbigny.
It is assumed that both the sample and anchor formed from istopic reservoirs with the same abundance of $\mnfivethree$, i.e., that $\mnfivethree$ is homogeneous among those two objects.
After this difference in formation time is determined using Mn-Cr systematics, the absolute Pb-Pb age of the anchor, $t_{\rm Pb,DOrbigny}$, is added, and a ``model" absolute age for the sample is determined. 
{\Steve The implicit goal of meteorite chronometry has been to obtain these absolute ages, and for that, use of individual anchors has been standard.}

{\Steve Here we argue that absolute ages are not the goal, and that recognition of this fact enables a move away from individual anchors, toward a more precise, statistical approach.}
{\Steve First it should be recognized that use of individual anchors introduces uncertainty to age determinations, especially since all model ages rely on Pb-Pb ages, which are uncertain, typically by $\pm 0.5$ Myr. 
In fact, there are two uncertain Pb-Pb ages: that of the anchor, whether that is the age of $t\!\!=\!\!0$,  usually taken to be the Pb-Pb age of CAIs \citep[e.g.,][]{ConnellyEtal2012}, or of an anchor like D'Orbigny; plus that of the sample.}
{\Steve Second, as discussed} in Paper I and above, this absolute age lacks meaning until it is put into a sequence of events in the early solar nebula, by determining the time of formation after $t\!\!=\!\!0$. 
This is done by subtracting the model age from the absolute age of the Solar System, $t_{\rm SS}$.
After all this, the time of formation after $t\!\!=\!\!0$ of the object, using Mn-Cr measurements, is calculated:
\begin{equation}
\Delta t_{53} = \tau_{53} \, \ln \left[ \frac{ (\mnratio)_{\rm DOrbigny} }{ (\mnratio)_0 } \right] + t_{\rm SS} - t_{\rm Pb,DOrbigny}. 
\label{eq:timetwo}
\end{equation} 
This is to be compared with the equivalent quantity in Equation~\ref{eq:timeone}.

It might seem that use of anchors avoids making assumptions about homogeneity of SLRs, in that it is not assumed that $\mnfivethree$ was homogeneous between the sample/anchor reservoir and the CAI-forming region.
However, it is still assumed that $\mnfivethree$ was homogeneous between the sample and the anchor, plus an {\it additional} assumption is made: that the Pb-Pb system in CAIs closed simultaneously with the Al-Mg system (or whatever system is used to define $t\!\!=\!\!0$). 
{\Steve In the end, deriving needed quantities like $\Delta t_{53}$ requires at least as many assumptions about homogeneity as just assuming} $\mnfivethree$ was homogeneous throughout the solar nebula and simply determining the correct value of $(\mnratio)_{\rm SS}$ to enter into Equation~\ref{eq:timeone}.

Effectively, {\Steve determining $(\mnratio)_{\rm SS}$} is just what the use of anchors does, as Equation~\ref{eq:timetwo} is equivalent to extrapolating backward in time from the anchor to define 
\begin{equation}
(\mnratio)_{\rm SS} = (\mnratio)_{\rm DOrbigny} \, \exp \left[ +(t_{\rm SS}-t_{\rm Pb,DOrbigny}) / \tau_{53} \right],
\label{eq:anchor}
\end{equation}
then using this value in Equation~\ref{eq:timeone} to find the time of formation of the sample.
Any time an anchor is used to infer when after $t\!\!=\!\!0$ a sample formed, it is equivalent to finding at least a model value for the initial abundance of an SLR in the solar system.
{\Steve The only difference is that the traditional approach using individual anchors determines $(\mnratio)_{\rm SS}$ using only a single meteorite at a time.}

Recognizing this mathematical equivalence allows a much more precise approach to chronometry, because in principle many anchors can be used simultaneously, to produce much less uncertain estimates of quantities such as $t_{\rm SS}$ and $(\mnratio)_{\rm SS}$.
In Paper I we showed that the Pb-Pb system {\Steve cannot be demonstrated to have achieved} isotopic closure in CAIs at the same time the Al-Mg system last did, so that Pb-Pb ages of CAIs are not likely to be a good estimate of $t_{\rm SS}$. 
Instead, we found that an age $t_{\rm SS} = 4568.42 \pm 0.24$ Myr minimized the discrepancies between times of formation determined by Al-Mg systematics, $\Delta t_{26}$, and times of formation determined by Pb-Pb ages, $\Delta t_{\rm Pb}$, for {\Steve seven} achondrites and four chondrules with simultaneous measurements. 
{\Steve Moreover, that analysis found that based on that value of $t_{\rm SS}$, concordancy was achieved in a statistically significant sense, justifying the assumption of homogeneity.}
Adopting a value for $t_{\rm SS}$, one can extrapolate backward from a sample like D'Orbigny to estimate $(\mnratio)_{\rm SS}$ in Equation~\ref{eq:anchor}. 
The estimates from several samples can be averaged together, producing a combined estimate for $(\mnratio)_{\rm SS}$ that is much more precise than could be achieved using one anchor alone.
{\Steve This approach, just like the traditional use of anchors, assumes homogeneity of SLRs; but using a statistical approach also allows this assumption to be tested rigorously.}

\subsection{Outline}

The goal of this paper {\Steve is to assess the homogeneity of other SLRs and use them to date meteorites.
In particular we aim} to determine the initial abundances of SLRs in the solar system, especially $(\mnratio)_{\rm SS}$ and $(\hfratio)_{\rm SS}$, but also $({}^{127}{\rm I}/{}^{129}{\rm I})_{\rm SS}$, $({}^{107}{\rm Pd}/{}^{108}{\rm Pd})_{\rm SS}$, and others. 
This assumes these SLRs were homogeneously distributed, an assumption we aim to test.
In Paper I we used statistical averages to find the value of $t_{\rm SS}$ that minimized the differences between the time of formation of a sample as inferred from Al-Mg systematics, $\Delta t_{26}$, and times of formation as determined from Pb-Pb ages, $\Delta t_{\rm Pb}$.
Here we will find the values of $t_{\rm SS}$, $(\mnratio)_{\rm SS}$, $\tau_{53}$, and $(\hfratio)_{\rm SS}$ that minimize the discrepancies between the times of formation of a sample as determined by Al-Mg, Mn-Cr and Hf-W, or Pb-Pb systematics.

In Paper I we restricted our attention to those samples that had both Al-Mg and Pb-Pb measurements.
In \S 2 we describe the 14 rapidly cooled achondrites we consider, that have Pb-Pb ages and at least one other age determination (Al-Mg, Mn-Cr, or Hf-W). 
We compile literature data to determine our best estimates of $(\alratio)_0$, $(\mnratio)_0$, $(\hfratio)_0$ and $t_{\rm Pb}$ for each.

In \S 3 we present a statistical approach we have developed.
We define a goodness-of-fit metric $\chisq$ for describing the degree to which the ages determined for various selected samples (achondrites) using the various isotopic systems (Al-Mg, Mn-Cr, Hf-W, {Pb-Pb) are concordant (based on the assumption of SLR homogeneity), and the threshold value of $\chisq$ for statistical significance. 
We also show how to optimize the input parameters $t_{\rm SS}$, $(\mnratio)_{\rm SS}$, $\tau_{53}$, and $(\hfratio)_{\rm SS}$, to minimize $\chisq$. 

In \S 4 we apply these statistical techniques to our dataset.
We find values for the four parameters $t_{\rm SS}$, $(\mnratio)_{\rm SS}$, $\tau_{53}$, and $(\hfratio)_{\rm SS}$, 
{\Steve that make the 37 available Al-Mg, Mn-Cr, Hf-W, and Pb-Pb formation times of the 14 achondrites concordant (excluding only the Hf-W formation time of NWA 4801).
The fit is statistically significant, with $\chisq = 1.09$ (33\% probability).}

In \S 5 we discuss the implications.
{\Steve The fact that a set of parameters makes all these formation times concordant fails to falsify, and instead strongly supports, the assumption that the SLRs were homogeneously distributed. 
Moreover, the}
values we derive for $(\mnratio)_{\rm SS}$ and $(\hfratio)_{\rm SS}$ compare favorably to values inferred from measurements of CAIs, and the mean-life we infer for $\mnfivethree$ is consistent with measurements, but our estimates are more precise. 
These results mean that one can use the assumed values of $(\mnratio)_{\rm SS}$ and $(\hfratio)_{\rm SS}$ to infer the time of formation of a sample without making reference to anchors.
The value we infer for $t_{\rm SS}$ is 1.1 Myr older than Pb-Pb ages of CAIs suggest, which we attribute (as in Paper I) to late resetting of the Pb-Pb system in CAIs.


In \S 6, we use our refined chronometry to refine the time of formation of Shallowater (used for I-Xe dating), and the initial ratios $({}^{60}{\rm Fe}/{}^{56}{\rm Fe})_{\rm SS}$ and  $({}^{107}{\rm Pd}/{}^{108}{\rm Pd})_{\rm SS}$, and others, to facilitate using these systems for radiometric dating. 
{\Steve We demonstrate that the concordancy extends to other systems, such as chondrules in CR chondrites, and chondrules in CB/CH chondrites.
We make suggestions for further tests of the model we present.}

We summarize our findings and draw conclusions in \S 7.

\section{Meteoritic Data}

\subsection{Sample Selection}

Our statistical technique of minimizing the discrepancies between the times of formation as determined by different isotopic systems (e.g., Al-Mg, Mn-Cr, Hf-W, Pb-Pb) presupposes that the different systems achieved isotopic closure simultaneously.
This drives us to select meteoritic samples that were melted (homogenizing the isotopes), rapidly cooled, and not obviously heated again or later metamorphosed (e.g., by shock).
Precise internal isochrons are required, so larger samples (e.g., achondrites) are preferred over smaller meteoritic components (e.g., chondrules or CAIs).

 Chondrules may seem like an excellent candidate for this type of analysis, and in some ways they are: 
the different isotopic systems have been measured in many individual chondrules, and chondrule textures indicate that they cooled and crystallized in a matter of only hours \citep{DeschEtal2012}.
However, over time spans of perhaps 2 Myr \citep{VilleneuveEtal2009}, chondrules appear to have experienced multiple transient heating events that raised them to different temperatures, including those near but not exceeding the solidus \citep{RuzickaEtal2008}.
As discussed in Paper I, at these temperatures and cooling rates it is possible to reset {\Steve the Pb-Pb chronometer without resetting the Al-Mg chronometer.
While the four chondrules considered were broadly concordant in their Al-Mg and Pb-Pb formation times, this may not necessarily be the case with chondrules overall.
A notable exception may be the chondrules produced in the impact associated with CB/CH chondrites, which were immediately swept up after formation.} 
We discuss chondrules in \S 5.3 and \S 5.4, but do not optimize the model to fit them.

Among achondrites, only a subset may be suitable for this analysis.
{\Steve In Paper I, we discussed the reasons why the isotopic systems should have closed simultaneously in the rapidly cooled [$\sim 300 \, {\rm K} \, {\rm hr}^{-1}$; \citep{Keil2012}] quenched angrites, including D'Orbigny, SAH 99555, and NWA 1670. 
The petrologically similar achondrites NWA 7325 and probably Asuka 881394, as well as NWA 2976 and NWA 6704 probably also cooled rapidly enough to pass through the closure temperatures of all isotopic systems essentially simultaneously.
This may not be the case for plutonic angrites.
Based on diffusion profiles of Ca in olivine, the plutonic angrite LEW 86010 is estimated to have cooled at about $300 \, {\rm K} \, {\rm yr}^{-1}$ \citep{McKayEtal1998}, about $10^4$ times more slowly than the quenched, or volcanic angrites. 
\citet{Keil2012} notes that these cooling rates are more consistent with near-surface dikes, shallow intrusions, or ponded lava flows, rather than true plutons. 
Still, the various isotopic systems with closure temperatures hundreds of K apart should have closed within only years of each other, essentially simultaneously.
It is also possible that a sample may see its isotopic systems achieve simultaneous isotopic closure, but then be reset by shock or metamorphism at a later time.
This may manifest itself as a resetting in some systems but not others, or in some just some rocks.
The plutonic angrite NWA 4801 may be one example of a disturbed sample \citep{IrvingKuehner2007,McKibbinEtal2015}.
}

Going forward, we restrict our attention to achondrites with U-corrected Pb-Pb ages and at least one other age from a different isotopic system. 
(We make an exception for Lewis Cliff 86010, whose Pb-Pb age is not U-corrected, as this is a much-studied angrite for which a reasonable guess to the $\uratio$ ratio can be made.
{\EDIT We also make an exception for NWA 1670 even though its uranium isotopes were not directly measured.})
For practical purposes, this usually ensures that a sample will have been measured in three systems, which provides a much more restrictive test of concordance. 
For systems formed more than about 5 Myr after $t\!\!=\!\!0$, after which $\altwosix$ is effectively extinct, it is almost the only way to ensure three ages for the same sample.

{Separate from the isotopic abundances associated with the decay of radionuclides are stable isotopic anomalies in bulk chondrites and achondrites, which provide important context for understanding their origins.
Isotopic evidence from $\epsilon^{50}{\rm Ti}$, $\epsilon^{54}{\rm Cr}$, and $\Delta^{17}{\rm O}$ isotopic ratios places the formation of meteorites in one of two reservoirs: the ``NC" reservoir, thought to be in the inner Solar System, in; or in the ``CC" reservoir, thought to be in the outer Solar System,\citep{TrinquierEtal2009,Warren2011,KruijerEtal2017}. 
All of the achondrites we include in our analysis are from the NC reservoir, except for the two recently dated achondrites NWA 2796 and NWA 6704, which derive from the CC reservoir \citep{SanbornEtal2019}.
}

In the following subsections we describe each achondrite used in our analysis, and the data used to derive $(\mnratio)_0$ and $(\hfratio)_0$ for each. 
The data used to derive $(\alratio)_0$ and Pb-Pb ages are discussed in Paper I.
{\Steve For all but two samples we apply the correction of 0.19 Myr advocated by \citet{TissotEtal2017}, to account for the empirical finding that pyroxenes are isotopically lighter than the whole rock.
{\EDIT We make exceptions for} the two achondrites from the CC isotopic reservoir, NWA 2976 and NWA 6704, on the basis that they formed from hydrous magmas in which U was more likely to be taken up entirely into pyroxene grains.}

\subsection{Quenched Angrites}

In contrast to plutonic angrites, which have nearly equilibrated minerals with little zoning, volcanic angrites have highly zoned mineral assemblages far from equilibrium.
{\EDIT \citep[See][]{TissotEtal2022}.}
`Quenched,' or `volcanic,' angrites are inferred to have cooled rapidly, at rates $\sim 300 \, {\rm K} \, {\rm hr}^{-1}$, after burial within the top meter or so of the surface \citep{Keil2012}.
If they escaped later resetting, they are likely to record simultaneous closure of the isotopic systems.

\subsubsection{D'Orbigny}

{\Steve D'Orbigny, described more fully in Paper I, is a quenched angrite that has long been considered an anchor in which the different isotopic systems likely closed simultaneously and has not been disturbed.}

As discussed in Paper I, we adopt for the initial $(\alratio)_0$ value for D'Orbigny the weighted mean of several values advocated by \citet{SanbornEtal2019}, $(3.93 \pm 0.39) \times 10^{-7}$.

For the $(\mnratio)_0$ ratio, we take the weighted mean of values determined by \citet{NyquistEtal2003}, \citet{GlavinEtal2004}, \citet{SugiuraEtal2005}, \citet{McKibbinEtal2015}, and \citet{KleineWadhwa2017}, to find $(3.233 \pm 0.033) \times 10^{-6}$, the value we adopt. 
An additional determination of $(3.20 \pm 0.21) \times 10^{-6}$ (consistent with our adopted value) was made by \citet{YinEtal2009}, but not in the refereed literature.

The ($\hfratio)_0$ ratio was determined by \citet{KleineEtal2012} to be $(7.15 \pm 0.17) \times 10^{-5}$.

As discussed in Paper I, for the Pb-Pb age we take the intercept of the Pb-Pb isochron derived by \citet{Amelin2008a}, and the weighted mean of the $\uratio$ values from \citet{BrenneckaWadhwa2012} and \citet{TissotEtal2017}, to find {\Steve $4563.24 \pm 0.21$} Myr. 

\subsubsection{SAH 99555}

{\Steve As described in Paper I}, SAH 99555 is a quenched angrite similar to D'Orbigny, with an unshocked, fine-grained texture composed of anorthite, Al-Ti-bearting hedenbergite, olivine and mm-sized vesicles \citep{Keil2012}.

As discussed in Paper I, we adopt for the initial $(\alratio)_0$ value for SAH 99555 the weighted mean of the values determined by \citet{SpivakBirndorfEtal2009} and \citet{SchillerEtal2015}, finding $(3.64 \pm 0.18) \times 10^{-7}$.

For the $(\mnratio)_0$ ratio, we take the weighted mean of values determined by  \citet{SugiuraEtal2005} and \citet{McKibbinEtal2015}, to find $(3.279 \pm 0.169) \times 10^{-6}$, the value we adopt. 

The ($\hfratio)_0$ ratio was determined by \citet{KleineEtal2012} to be $(6.87 \pm 0.15) \times 10^{-5}$.

As discussed in Paper I, for the Pb-Pb age we take the average of the intercepts of the Pb-Pb isochrons derived by \citet{Amelin2008b} and \citet{ConnellyEtal2008}, and the $\uratio$ value determined by \citet{BrenneckaWadhwa2012} and \citet{TissotEtal2017}, to find 
{\Steve $4563.51 \pm 0.24$ Myr}. 

\subsubsection{NWA 1670}

NWA 1670, as described in Paper I, is a quenched angrite with a porphyritic texture including large olivine megacrysts in a fine-grained matrix of olivine, pyroxene, kirsch-steinite and anorthite, as well as other accessory minerals \citep{Keil2012}.
These indicate rapid cooling at $\sim 300 \, {\rm K} \, {\rm hr}^{-1}$ \citep{MikouchiEtal2003}.

For NWA 1670, we adopt the $(\alratio)_0$ ratio determined by \citet{SchillerEtal2015}, $(5.92 \pm 0.59) \times 10^{-7}$.

For $(\mnratio)_0$, we adopt the value determined by \citet{SugiuraEtal2005}, $(2.85 \pm 0.92) \times 10^{-6}$.

We are not aware of an determination of $(\hfratio)_0$ for NWA 1670.

As described fully in Paper I \citep{DeschEtal2023a}, we have reanalyzed the Pb-Pb isochron of \citet{SchillerEtal2015} to determine a Pb-Pb age {\Steve $4564.02 \pm 0.66$ Myr}.

\subsubsection{NWA 1296}
{\DRD 
NWA 1296 is a quenched angrite with a bulk composition similar to that of D'Orbigny and Sahara 99555 \citep{JambonEtal2004,TissotEtal2022}. NWA 1296 has fine-grained texture consisting of primarily of dentritic olivine, anorthite and Al-Fe diopside-hedenbergite pyroxenes \citep{JambonEtal2004}. The texture and grain sizes are consistent with formation through rapid crystallization.}
{\Steve 
We are not aware of Al-Mg or Mn-Cr measurements for this achondrite.

We adopt $(\hfratio)_0 = (7.01 \pm 0.28) \times 10^{-5}$ \citep{KleineEtal2012}.

Its Pb-Pb age was determined by \citet{AmelinIrving2011metsoc} to be $4564.20 \pm 0.45$ Myr, based on an assumed $\uratio = 137.88$.
We are not aware of any measurements in the refereed literature, or indeed of any direct measurements of its uranium isotopes, but we retain this age as a test of the model. 
The most severe test comes from assuming the youngest plausible Pb-Pb age, i.e., by assuming the minimum $\uratio$ value.
We adopt the value $\uratio = 137.786 \pm 0.013$ from NWA 1670 \citep{SchillerEtal2015}, which implies a correction -0.99 Myr.
As with the other NC achondrites, we assume corrections like that applied to NWA 1670 are based on measurements of pyroxene grains, which are isotopically lighter than whole-rock measurements, and apply an additional correction -0.19 Myr.
This yields an age $4563.02 \pm 0.45$ Myr. 
}

\subsection{Plutonic Angrites}

In contrast to quenched angrites, plutonic angrites appear to have achieved equilibrium, cooling much more slowly than volcanic angrites, at rates $\sim 300 \, {\rm K} \, {\rm yr}^{-1}$, indicating burial at depths of tens of meters \citep{Keil2012}, 
This cooling rate is still rapid enough for their isotopic systems to have achieved closure simultaneously, but many plutonic angrites appear to have been later disturbed or metamorphosed, which may affect different systems differently.

\subsubsection{LEW 86010}

LEW 86010 is an unshocked plutonic angrite with granular texture, with grains 0.6-1.2 mm across, composed of anorthite, Al-Ti-bearing diopside, and calcic olivine, with some kirschsteinite. 
It is thought to have cooled in thousands of years or less, based on zoning in pyroxene and exsolution lamellae in olivine \citep{McKayEtal1998,Keil2012,McKibbinEtal2015}.

We are not aware of determinations of ($\alratio)_0$ for LEW 86010 or any of the plutonic angrites.
This is unsurprising, as the other systems suggest they formed roughly 10 Myr after $t\!\!=\!\!0$, when $\altwosix$ would have been effectively extinct. 

For the $(\mnratio)_0$, we adopt the weighted mean of values determined by \citet{LugmairShukolyukov1998} and \citet{NyquistEtal1994}, to find $(1.345 \pm 0.049) \times 10^{-6}$.

The ($\hfratio)_0$ ratio was determined by \citet{KleineEtal2012} to be $(4.80 \pm 0.42) \times 10^{-5}$.

{\Steve
The Pb-Pb age of LEW 86010 was determined by \citet{Amelin2008a} to be 
$4558.55 \pm 0.15$ Myr, 
but without using a measurement of the $\uratio$ in the sample, instead assuming 137.88. 
While LEW 86010 is important for its previous use as an anchor, its small size (6.9 g) has precluded precise measurement of its $\uratio$,
although \citet{LugmairGaler1992} found the ${}^{238}{\rm U}/{}^{235}{\rm U}$ ratio was about $(1.1 \pm 1.7) \permil$ lighter than 137.88 in its pyroxenes (and the whole rock is isotopically heavier, in line with other achondrites).
Correcting for this would lower the age of LEW 86010 by $1.6 \pm 2.5$ Myr. 
Adopting the value $\uratio = 137.786$ apparently common to plutonic angrites \citep{TissotEtal2017}, we estimate an age correction of -0.99 Myr, but with considerable {\EDIT uncertainty}.
For these reasons we cannot determine the Pb-Pb age of LEW 86010 with certainty, but based on its previous use as an anchor, we include it in our analysis.
}

\subsubsection{NWA 4590}

NWA 4590 is a coarse-grained igneous cumulate rock with Al-Ti-rich clinopyroxene, anorthite, Ca-rich olivine with kirchsteinite exsolution, ulv\"{o}-spinel, plus merrillite and silico-phosphate. 
As with LEW 86010, it is thought to have cooled over only thousands of years, based on zoning in pyroxene and exsolution lamellae in olivine \citep{McKibbinEtal2015}.
Pb-Pb dating has been applied to the silicates and silico-phosphates, and the Pb-Pb ages found to differ by $0.55 \pm 0.29$ Myr; based on the differences in closure temperatures, a slow cooling rate $540 \pm 290 \, {\rm K} \, {\rm Myr}^{-1}$ was inferred \citep{AmelinEtal2011}.
However, other petrologic evidence suggests that instead the Pb-Pb system in the phosphates was reset by a later reheating event \citep{McKibbinEtal2015}.

For the $(\mnratio)_0$ value, we adopt the value determined by \citet{McKibbinEtal2015}, $(0.85 \pm 0.40) \times 10^{-6}$.
We note that \citet{YinEtal2009}, in unrefereed work, found a similar value, $(1.01 \pm 0.12) \times 10^{-6}$. 

The ($\hfratio)_0$ ratio was determined by \citet{KleineEtal2012} to be $(4.63 \pm 0.17) \times 10^{-5}$.

The Pb-Pb age was determined to be $4557.81 \pm 0.37$ Myr by \citet{BrenneckaWadhwa2012}, who applied a measured uranium correction to the Pb-Pb isochrons measured by \citet{AmelinIrving2007} and \citet{AmelinEtal2011}. 
A more refined uranium correction was applied by 
\citet{TissotEtal2017}, who determined a Pb-Pb age of $4557.76 \pm 0.38$ Myr.
{\Steve After applying a 0.19 Myr correction, we adopt a value $4557.57 \pm 0.38$ Myr}.

\subsubsection{NWA 4801}

NWA 4801 has a granular, cumulate texture, with grain sizes 0.1 - 1.2 mm, described by \citet{IrvingKuehner2007} as being an annealed breccia formed originally by disruption of a very coarse-grained plutonic protolith.
It consists primarily of Al-Ti-bearing diopside and anorthite. 
{\DRD \citet{McKibbinEtal2015} explain the lack of chemical variation in pyroxene and olivine, along with the textural features, as petrologic evidence for slow cooling, but also mention a later stage of high-temperature annealing could have occured as was described by \citet{IrvingKuehner2007}.}

For the $(\mnratio)_0$ value, although it is the only value in the refereed literature, we consider the value determined by \citet{McKibbinEtal2015}, $(0.13 \pm 1.1) \times 10^{-6}$, to be too imprecise.
{\EDIT We consider the value reported by \citet{ShukolyukovEtal2009}, $(0.96 \pm 0.04) \times 10^{-6}$, to be overly precise.
We adopt the value from the abstract by} \citet{YinEtal2009}, $(0.959 \pm 0.040) \times 10^{-6}$. 

The ($\hfratio)_0$ ratio was determined by \citet{KleineEtal2012} to be $(4.52 \pm 0.16) \times 10^{-5}$.

The Pb-Pb age was determined to be $4557.01 \pm 0.27$ Myr by \citet{BrenneckaWadhwa2012}, who applied a measured uranium correction to the Pb-Pb isochron measured by \citet{Amelin2008a}. 
A more refined uranium correction was applied by 
\citet{TissotEtal2017}, who determined a Pb-Pb age of $4556.82 \pm 0.28$ Myr.
{\EDIT A value of $4556.8 \pm 0.2$ Myr was reported by \citet{ConnellyBizzarro2016}. 
We take the weighted mean of these to find $4556.91 \pm 0.21$ Myr.}
{\Steve After correcting this by 0.19 Myr, we adopt $4556.72 \pm 0.21$ Myr.}

\subsubsection{Angra dos Reis}

Angra dos Reis (AdoR) is a porphyritic igneous rock composed of pyroxene, Al-Ti-bearing diopside, calcic olivine, and other minerals.
Despite being the namesake of angrites, AdoR differs from other angrites in its geochemical composition, as it is nearly mono-mineralic with ${>}$90 vol{\%} pyroxene \citep{PrinzEtal1977}.

For the $(\mnratio)_0$ value, \citet{LugmairShukolyukov1998} combined data for AdoR with that for LEW 86010 and found they were consistent with the same isochron, 
{\Steve 
with slope $(1.25 \pm 0.07) \times 10^{-6}$. 
However, it is unlikely that AdoR and LEW 86010 formed {\EDIT in the same magmatic system}, and therefore unlikely that they should close at the same time.
Instead, we take the two data points 
and derive a slope $(\mnratio)_0 = (1.10 \pm 0.40) \times 10^{-6}$.
}

The Pb-Pb age was determined to be $4556.60 \pm 0.26$ Myr by \citet{BrenneckaWadhwa2012}, who applied a measured uranium correction to the Pb-Pb isochrons measured by \citet{Amelin2008a}. 
A more refined uranium correction was applied by 
\citet{TissotEtal2017}, who determined a Pb-Pb age of $4556.45 \pm 0.29$ Myr.
{\Steve After correcting this by 0.19 Myr, we adopt $4556.26 \pm 0.29$ Myr.}

\subsubsection{NWA 2999 and Paired Meteorites}

Containing both coarse and fine grained lithologies, NWA 2999 (and numerous pairings, {\EDIT including NWA 4931, NWA 6291, and at least six others}) has been described as a plutonic angrite \citep{Keil2012} as well as an annealed breccia (\citealt{IrvingKuehner2007}). 
Due to the high abundance and siderophile element contents of metal, an exogenous impactor has been suggested to mix materials of carbonaceous chondrite origin from the CC reservoir \citep{GellissenEtal2007,HumayunEtal2007,RichesEtal2012}, even though the W isotope composition would seem to preclude this \citep{KleineEtal2012}.
NWA 2999 shows signs of terrestrial alteration in the presence of iron replacement minerals (goethite and magnetite) as well as light rare earth element (LREE) enrichments and Ce anomalies in olivine \citep{SanbornWadhwa2021}. 
The trace element data present by \citet{SanbornWadhwa2021} shows that NWA 2999 has a composition more closely related to volcanic angrites and implies that magmatic activity of the volcanic angrite source reservoir continued for millions of years.
Consensus on the petrogenesis of NWA 2999 has not been reached and in many ways it is a confusing member of the angrite group.

For the $(\mnratio)_0$ ratio, the only value we could find was in the unrefereed abstract by \citet{ShukolyukovLugmair2008}, $(1.28 \pm 0.23) \times 10^{-6}$.
{\Steve 
We do not consider this a reliable Mn-Cr formation time of NWA 2999.}

The ($\hfratio)_0$ ratio was determined by \citet{KleineEtal2012} to be $(5.43 \pm 0.34) \times 10^{-5}$.

For the Pb-Pb age, we take the value $4560.74 \pm 0.47$ Myr determined by \citet{BrenneckaWadhwa2012}, based on a Pb-Pb isochron by \citet{AmelinIrving2007}.
{\Steve After correcting this by 0.19 Myr, we adopt $4560.55 \pm 0.47$ Myr.}

\subsection{Other NC Achondrites} 

\subsubsection{Asuka 881394}

 {\DRD Asuka 881394 is a eucrite-like achondrite with a coarse-grained igneous texture with near equal amounts of anorthite and pyroxene. }As discussed in Paper I, we adopt for the initial $(\alratio)_0$ value for Asuka 881394 the weighted mean of the values determined by \citet{NyquistEtal2003}, \citet{WadhwaEtal2009}, and \citet{WimpennyEtal2019}, finding $(13.071 \pm 0.55) \times 10^{-7}$.

For the $(\mnratio)_0$ ratio, we take the weighted mean of values determined by  \citet{NyquistEtal2003} and \citet{WimpennyEtal2019}, to find $(3.863 \pm 0.228) \times 10^{-6}$, the value we adopt. 

We are not aware of a determination of $(\hfratio)_0$ for Asuka 881394. 

The Pb-Pb age of Asuka 881394 was determined by \citet{WadhwaEtal2009} to be $4566.5 \pm 0.2$ {\Steve (not U-corrected)}, and by  \citep{WimpennyEtal2019} to be $4564.95 \pm 0.53$ Myr, the value we take. 
{\Steve After correcting this by 0.19 Myr, we adopt $4564.76 \pm 0.53$ Myr.}

\subsubsection{Ibitira}

Ibitira is an unbrecciated basaltic rock with abundant vesicles and a fine-grained texture.
It is compared to eucrites, but is distinct.
Its plagioclase is mostly calcic, due to depletion in alkali elements, and its pyroxenes have high Fe/Mn ratios in comparison to typical basaltic eucrites.
\citet{Mittlefehldt2005} argued these indicate formation on a distinct parent asteroid, a conclusion corroborated by the finding that the $\Delta^{17}{\rm O}$ oxygen isotope composition is $16-21{\sigma}$ above the HED mean values \citep{ScottEtal2009}.
The vesicles in Ibitira suggest a volcanic origin, although its seems to have formed at the same time as plutonic angrites.
It is likely a volcanic achondrite, but we treat it separately. 

We are not aware of determinations of $(\alratio)_0$ or $(\hfratio)_0$ for Ibitira.

For the $(\mnratio)_0$ value, we adopt the value of \citet{LugmairShukolyukov1998}, $(1.06 \pm 0.50) \times 10^{-6}$.
We note that a similar value was reported in the unrefereed abstract of \citet{YinEtal2009}.

We take the Pb-Pb age of $4556.75 \pm 0.57$ of \citet{IizukaEtal2014} for Ibitira.
{\Steve After correcting this by 0.19 Myr, we adopt $4556.56 \pm 0.57$ Myr.}

\subsubsection{NWA 7325}

{\DRD As described in paper I, NWA 7325 is an ungrouped achondrite with a medium-grained cumulate texture consisting of Mg-rich olivine, Cr-bearing diopside and Ca-rich plagioclase \citep{GoodrichEtal2017}.}

As in Paper I, we adopt the value $(\alratio)_0 = (3.03 \pm 0.14) \times 10^{-7}$ \citep{KoefoedEtal2016} for NWA 7325.

We are not aware of determinations of $(\mnratio)_0$ or $(\hfratio)_0$ for NWA 7325.

As in Paper I, we adopt {\Steve $4563.7 \pm 1.7$ Myr} \citep{KoefoedEtal2016} for the Pb-Pb age of NWA 7325.

\subsection{CC Achondrites}

\subsubsection{NWA 2976}

{\DRD
As described in Paper I, NWA 2976 (paired with NWA 011) is an unshocked, unbrecciated ungrouped achondrite with coarse grained pigeonite surrounded by fine-grained, recrystallized plagioclase with well-developed 120$^{\circ}$ triple junctions \citep{YamaguchiEtal2002}.}

As in Paper I, we adopt {\Steve a weighted mean of the values from \citet{BouvierEtal2011b} and \citet{SchillerEtal2010}, $(\alratio)_0 = (4.05 \pm 0.15) \times 10^{-7}$.}

We are not aware of a determination of $(\mnratio)_0$ or $(\hfratio)_0$ for NWA 2976.

As in Paper I, we adopt $4563.16 \pm 0.57$ Myr for the Pb-Pb age of NWA 2976.

\subsubsection{NWA 6704} 

{\DRD As described in Paper I, NWA 6704 (paired with NWA 6693) is an unshocked, ungrouped achondrite with a medium grained texture comprised of low-Ca pyroxene along with Ni-rich olivine and sodic plagioclase \citep{HibiyaEtal2019}.}
As in Paper I, we adopt $(\alratio)_0 = (3.15 \pm 0.38) \times 10^{-7}$ value for NWA 6704 \citep{SanbornEtal2019}.

We adopt the value $(\mnratio)_0 = (2.59 \pm 0.34) \times 10^{-6}$ for NWA 6704 \citep{SanbornEtal2019}.

As in Paper I, we adopt $4562.76 \pm 0.26$ Myr for the Pb-Pb age of NWA 6704 \citep{AmelinEtal2019}.

\subsection{Summary of Achondrite Data}

In Table~\ref{table:one} we compile all of the above data, which comprise 38 values of either 
$(\alratio)_0$, $(\mnratio)_0$, $(\hfratio)_0$, or U-corrected Pb-Pb ages $t_{\rm Pb}$, across 14 achondrites.
In general, we only use values from the refereed literature; but in some cases, if other data were not available, we used data from abstracts (as noted in the caption).  
Where multiple measurements are available for a single achondrite, we have taken a weighted average. 
With knowledge of key quantities such as the $\mnfivethree$ half-life, and the Pb-Pb age of samples formed at $t\!\!=\!\!0$, $t_{\rm SS}$, these quantities can be converted into times of formation after $t\!\!=\!\!0$, next.

$\!\!\!\!\!\!\!\!\!\!\!\!\!\!\!\!\!\!\!\!\!\!\!\!\!\!\!\!\!\!\!\!$
\begin{table}[h!]
\noindent
\begin{minipage}{6.5in}
\footnotesize
\caption{$(\alratio)_0$, $(\mnratio)_0$, $(\hfratio)_0$, and U-corrected Pb-Pb ages of 13 achondrites.
\label{table:one}}
\begin{tabular}{l|rrc|rrc|ccc|rrc}
{\bf Achondrite} & 
$\left( \frac{{}^{26}{\rm Al}}{{}^{27}{\rm Al}} \right)_0$ & $2\sigma$ & $\dagger$ & 
$\left( \frac{{}^{53}{\rm Mn}}{{}^{55}{\rm Mn}} \right)_0$ & $2\sigma$ & $\dagger$ & 
$\left( \frac{{}^{182}{\rm Hf}}{{}^{180}{\rm Hf}} \right)_0$& $2\sigma$ & $\dagger$ & 
Pb-Pb          & $2\sigma$ & $\dagger$ \\
\hline
{\bf D'Orbigny} & {\bf 3.93} & {\bf 0.39} & {\it *} & {\bf 3.23} & {\bf 0.03} & {\it *} & {\bf 7.15} & {\bf 0.17} & {\it i} & {\bf 4563.24} & {\bf 0.21} & {\it j} \\
$\,$  & 5.06  & 0.92        & {\it a} &
  2.83       & 0.25        & {\it e} & 
             &             &         & 
             &             &         \\
$\,$  & 3.98       & 0.15        & {\it b} &
  3.24       & 0.04        & {\it f} & 
             &             &         & 
             &             &         \\ 
$\,$ & 3.97       & 0.21        & {\it c} &
  2.84       & 0.24        & {\it g} & 
             &             &         & 
             &             &         \\ 
$\,$ & 3.93       & 0.39        & {\it d} &
  3.54       & 0.18        & {\it h} & 
             &             &         & 
             &             &         \\ 
$\,$ & $\,$       & $\,$        & $\,$ &
  3.23       & 0.07        & {\it c} & 
             &             &         & 
             &             &         \\ 
\hline
 {\bf SAH 99555} & {\bf 3.65} & {\bf 0.18} & {\it *} & {\bf 3.28} & {\bf 0.17} & {\it *} & {\bf 6.87} & {\bf 0.15} & {\it i} & {\bf 4563.51} & {\bf 0.24} & {\it k} \\
$\,$ & 5.13       & 1.90        & {\it a} &
  2.82       & 0.37        & {\it g} & 
             &             &         & 
             &             &         \\ 
$\,$ & 3.64       & 0.18        & {\it b} &
  3.40       & 0.19        & {\it h} & 
             &             &         & 
             &             &         \\ 
\hline
{\bf NWA 1670} & {\bf 5.92} & {\bf 0.59} & {\it b} & {\bf 2.85} & {\bf 0.92} & {\it g} & & & & {\bf 4564.02} & {\bf 0.66} & {\it l} \\
\hline
{\bf NWA 1296} & {} & {} & {} & {} & {} & {} & {\bf 7.01} & {\bf 0.28} & {\it i} & {\bf 4563.02} & {\bf 0.45} & {\it m} \\
\hline
{\bf LEW 86010} & & & & {\bf 1.35} & {\bf 0.05} & {\it *} & {\bf 4.80} & {\bf 0.42} & {\it i} & {} & {} & {} \\
$\,$ &            &             &         &
  1.44       & 0.07        & {\it n} & 
             &             &         & 
             &             &         \\ 
$\,$ &            &             &         &
  1.25       & 0.07        & {\it o} & 
             &             &         & 
             &             &         \\ 
\hline
{\bf NWA 4590} & & & & {\bf 0.85} & {\bf 0.40} & {\it h} & {\bf 4.63} & {\bf 0.17} & {\it i} & {\bf 4557.57} & {\bf 0.38} & {\it *} \\
$\,$ & & & &  &  &  &  & &  & {4557.81} & {0.37} & {\it p} \\
$\,$ & & & &  &  &  &  & &  & {4557.76} & {0.38} & {\it q} \\
\hline
 {\bf NWA 4801} & & & & {\bf 0.96} & {\bf 0.04} & {\it *} & {\bf 4.52} & {\bf 0.16} & {\it i} & {\bf 4556.63} & {\bf 0.28} & {\it *} \\
$\,$ &            &             &         &
  0.919      & 0.295       & {\it r} & 
             &             &         & 
  4557.01    & 0.27        & {\it t} \\ 
$\,$ &            &             &         &
  0.96       & 0.04        & {\it s} & 
             &             &         & 
  4556.82    & 0.28        & {\it q} \\ 
\hline
{\bf Angra} & & & & {\bf 1.10} & {\bf 0.40} & {\it n} & {\bf 4.02} & {\bf 0.24} & {\it i} & {\bf 4556.26} & {\bf 0.29} & {\it *} \\
{\bf dos Reis} &            &             &         &
             &             &         & 
             &             &         & 
  4556.60    & 0.26         & {\it u} \\ 
$\,$ &            &             &         &
             &             &         & 
             &             &         & 
  4556.45    & 0.29  & {\it q} \\
\hline
{\bf NWA 2999} & & & & {\bf 1.28} & {\bf 0.23} & {\it v} & {\bf 5.43} & {\bf 0.34} & {\it i} & {\bf 4560.55} & {\bf  0.47} & {\it t} \\
\hline
{\bf Asuka 881394} & {\bf 13.07} & {\bf 0.56} & {\it *} & {\bf 3.86} & {\bf 0.23} & {\it *} & & & & {\bf 4564.76} & {\bf 0.53} & {\it *} \\ 
$\,$ & 11.8       & 1.4         & {\it e} &
  4.6        & 1.7         & {\it e} & 
             &             &         & 
  4566.5     & 0.2         & {\it w} \\
$\,$ & 12.8       & 0.7         & {\it w} &
  3.85       & 0.23        & {\it x} & 
             &             &         & 
  4564.95    & 0.53        & {\it x} \\
$\,$ & 14.8       & 1.2         & {\it x} &
             &             &         & 
             &             &         & 
             &             &         \\
\hline
{\bf Ibitira}  &            &            &         & {\bf 1.06} & {\bf 0.50} & {\it n} & & & & {\bf 4556.56} & {\bf 0.57} & {\it y} \\ 
\hline
{\bf NWA 7325} & {\bf 3.03} & {\bf 0.14} & {\it z} & & & & & & & {\bf 4563.7}  & {\bf 1.7}  & {\it z} \\
\hline
{\bf NWA 2976} & {\bf 4.05} & {\bf 0.15} & {\it *} & & & & & & & {\bf 4563.16} & {\bf 0.57} & {\it aa} \\
$\,$ & 4.91       & 0.46        & {\it aa} &
             &             &         & 
             &             &         & 
             &             &         \\
$\,$ & 3.94       & 0.16        & {\it bb} &
             &             &         & 
             &             &         & 
             &             &         \\
\hline
{\bf NWA 6704} & {\bf 3.15} & {\bf 0.38} & {\it d} & {\bf 2.59} & {\bf 0.34} & {\it d} & & & & {\bf 4562.76} & {\bf 0.26} & {\it cc} \\ 
\hline  
\end{tabular}
\footnotesize{
$(\alratio)_0$ and $2\sigma$ uncertainties in units of $10^{-7}$; $(\mnratio)_0$ and $2\sigma$ uncertainties in units of $10^{-6}$; $(\hfratio)_0$ and $2\sigma$ uncertainties in units of $10^{-5}$; Pb-Pb ages and uncertainties in units of Myr. Adopted values in bold type.  {\it *} denotes a weighted average of literature data;
$\dagger$ References: 
 {\it a.} \citet{SpivakBirndorfEtal2009};
 {\it b.} \citet{SchillerEtal2015}; 
 {\it c.} \citet{KleineWadhwa2017};
 {\it d.} \citet{SanbornEtal2019};
 {\it e.} \citet{NyquistEtal2003};
 {\it f.} \citet{GlavinEtal2004};
 {\it g.} \citet{SugiuraEtal2005};
 {\it h.} \citet{McKibbinEtal2015};
 {\it i.} \citet{KleineEtal2012}; 
 {\it j.} Pb-Pb isochron from \citet{Amelin2008a}, uranium correction from \citet{BrenneckaWadhwa2012} and \cite{TissotEtal2017} [see \citealt{DeschEtal2023a}];
 {\it k.} Pb-Pb isochron from \citet{Amelin2008b} and \citet{ConnellyEtal2008}, uranium correction from \citet{TissotEtal2017} and \citet{ConnellyEtal2012} [see \citealt{DeschEtal2023a}]; 
 {\it l.} \citet{SchillerEtal2015}, as corrected by \citet{DeschEtal2023a}; 
 {\it m.} \citet{AmelinIrving2011metsoc}, with corrections described in text;
 {\it n.} \citet{NyquistEtal1994};
 {\it o.} \citet{LugmairShukolyukov1998};
 {\it p.}
 \citet{BrenneckaWadhwa2012}, based on Pb-Pb isochron of \citet{AmelinIrving2007} (abstract) and \citet{AmelinEtal2011} (abstract).
 {\it q.} 
 \citet{TissotEtal2017};
 {\it r.} \citet{YinEtal2009} (abstract);
 {\it s.} \citet{ShukolyukovEtal2009} (abstract);
 {\it t.} \citet{BrenneckaWadhwa2012}, based on Pb-Pb isochron by \citet{AmelinIrving2007};
 {\it u.} \citet{BrenneckaWadhwa2012}, based on Pb-Pb isochron from \citet{Amelin2008a};
 {\it v.} \citet{ShukolyukovLugmair2008} (abstract);
 {\it w.}  \citet{WadhwaEtal2009};
 {\it x.} \citet{WimpennyEtal2019};
 {\it y.} \citet{IizukaEtal2014}; 
 {\it z.} \citet{KoefoedEtal2016}; 
 {\it aa.} \citet{SchillerEtal2010};
 {\it bb.} \citet{BouvierEtal2011b}; 
 {\it cc.} 
 \citet{AmelinEtal2019}.
 }
\end{minipage}
\end{table}
\section{Statistical Chronometry Methods}

We advocate analysis of the above samples through a statistical approach.
This builds on {\Steve similar approaches: for example, that of} \citet{NyquistEtal2009}, who correlated Mn-Cr and Al-Mg ages, and estimated $t_{\rm SS}$, among other quantities; \citet{TissotEtal2017}, who averaged several samples together to better estimate $(\mnratio)_{\rm SS}$; \citet{SanbornEtal2019}, who correlated several samples together to better constrain the $\mnfivethree$ half-life;
{\Steve and \citet{PirallaEtal2023}, who correlated Al-Mg and Pb-Pb ages to find the age of the Solar System.}
Our approach, which we call ``statistical chronometry," {\Steve is more comprehensive, and} has the goal of fitting all the systems simultaneously, to derive the values that make all the isotopic systems concordant, in the achondrites for which they are most likely to be concordant;
{\Steve and is unique in employing rigorous methods to statistically evaluate the goodness of that fit}.

\subsection{Goodness-of-fit Metric}

The goal of our statistical approach is to find the optimal values of: $t_{\rm SS}$, the Pb-Pb age of samples that closed at $t\!\!=\!\!0$, defined to be the time at which $(\alratio) = (\alratio)_{\rm SS} = 5.23 \times 10^{-5}$; $(\mnratio)_{\rm SS}$ and $(\hfratio)_{\rm SS}$, the abundances of $\mnfivethree$ and $\hfoneeighttwo$ at $t\!\!=\!\!0$; and the half-lives of $\mnfivethree$, $\altwosix$, and $\hfoneeighttwo$.
The optimal values are those that minimize for each achondrite the differences between the times of formation after $t\!\!=\!\!0$ as determined by each isotopic system ($\Delta t_{26}$, $\Delta t_{53}$, $\Delta t_{182}$, and $\Delta t_{\rm Pb}$), and the ``true" or best estimate of its time of formation, $\Delta t$. 
More specifically, if we assume the 
initial solar system abundance of $\altwosix$ is a fixed quantity for the purpose of the calculation, the times of formation one would infer from the Al-Mg system would be
\begin{equation}
\Delta t_{26} = \tau_{26} \, \ln \left[ \frac{ (\alratio)_{\rm SS} }{ R_{26} } \right],
\end{equation}
with ($2\sigma$) uncertainty $\sigma_{\Delta t26}$, where 
\begin{equation}
\sigma_{\Delta t26}^2 = \tau_{26}^2 \, \left( \frac{ \sigma_{R26} }{ R_{26} } \right)^2 + \left( \Delta t_{26} \right)^2 \, \left( \frac{\sigma_{\tau26}}{\tau_{26}} \right)^2.
\end{equation}
Here $R_{26} = (\alratio)_{0}$, and $\sigma_{R26}$ is the ($2\sigma$) uncertainty in $(\alratio)_{0}$, and for the purposes of calculating the goodness-of-fit metric we will assume that any mean-life (e.g., $\tau_{26}$) is a fixed input parameter, in which case the measurement uncertainty $\sigma_{\tau26}$ is irrelevant. 
Likewise, 
\begin{equation}
\Delta t_{53} = \tau_{53} \, \ln \left[ \frac{ (\mnratio)_{\rm SS} }{ R_{53} } \right]
\end{equation}
and 
\begin{equation}
\sigma_{\Delta t53} = \tau_{53} \, \frac{ \sigma_{R53} }{ R_{53} },
\end{equation}
and 
\begin{equation}
\Delta t_{182} = \tau_{182} \, \ln \left[ \frac{ (\hfratio)_{\rm SS} }{ R_{182} } \right]
\end{equation}
and 
\begin{equation}
\sigma_{\Delta t182} = \tau_{182} \, \frac{ \sigma_{R182} }{ R_{182} }.
\end{equation}
In a similar vein, assuming $t_{\rm SS}$ is a fixed quantity for the purposes of the calculation, 
\begin{equation}
\Delta t_{\rm Pb} = t_{\rm SS} - t_{\rm Pb},
\end{equation}
where $t_{\rm Pb}$ is the Pb-Pb age of the sample, and the ($2\sigma$) uncertainty is
\begin{equation}
\sigma_{\Delta tPb} = \sigma_{\rm tPb},
\end{equation}
where $\sigma_{\rm tPb}$ is the ($2\sigma$) uncertainty in the Pb-Pb age.

If all of the isotopic systems achieved closure at the same time {\Steve in an achondrite}, then there is one time of formation, $\Delta t$, and each of these isotopic systems is providing an estimate or ``measurement" of $\Delta t$. 
The best estimate of the time of formation is then {\Steve the weighted mean}:
\begin{equation}
\Delta t = \left[ \frac{ \Delta t_{26} }{ \sigma_{\Delta t26}^2 } + \frac{ \Delta t_{53} }{ \sigma_{\Delta t53}^2 } + \frac{ \Delta t_{182} }{ \sigma_{\Delta t182}^2 } + \frac{ \Delta t_{\rm Pb} }{ \sigma_{\Delta tPb}^2 } \right] \div
\left[ \frac{ 1 }{ \sigma_{\Delta t26}^2 } + \frac{ 1 }{ \sigma_{\Delta t53}^2 } + \frac{ 1 }{ \sigma_{\Delta t182}^2 } + \frac{ 1 }{ \sigma_{\Delta tPb}^2 } \right].
\label{eq:deltat}
\end{equation}
Here it is understood that for samples for which no age information exists, $\sigma_{\Delta t}$ is effectively infinite.

Based on this best estimate, we can define a goodness-of-fit parameter describing how well the isotopic systems combined fit the best-fit $\Delta t_i$ for one achondrite indexed by $i$; when summed over all achondrites from $i=1$ to $i=A$, we derive a global goodness-of-fit parameter:
\[
\!\!\!\!\!\!\!\!\!\!\!\!\!\!\!\!\!\!\!\!\!\!\!
\!\!\!\!\!\!\!\!\!\!\!\!\!\!\!\!\!\!\!\!\!\!\!
\!\!\!\!\!\!\!\!\!\!\!\!\!\!\!\!\!\!\!\!\!\!\!
\!\!\!\!\!\!\!\!\!\!\!\!\!\!\!\!\!\!\!\!\!\!\!
\chi_{\nu}^{2} = \frac{1}{N-M} \, \sum_{i=1}^{A} \left[ \frac{ \left( \Delta t_{26,i} - \Delta t_{i} \right)^2 }{ \sigma_{\Delta t26,i}^2 } \right.
\]
\begin{equation}
 \left. +\frac{ \left( \Delta t_{53,i} - \Delta t_{i} \right)^2 }{ \sigma_{\Delta t53,i}^2 } +\frac{ \left( \Delta t_{182,i} - \Delta t_{i} \right)^2 }{ \sigma_{\Delta t182,i}^2 } +\frac{ \left( \Delta t_{{\rm Pb},i} - \Delta t_{i} \right)^2 }{ \sigma_{\Delta t{\rm Pb},i}^2 } \right].
 \label{eq:chisq}
\end{equation}
Here $N$ ($\leq 37)$ is the number of ages across all $A$ ($\leq 14$) achondrites that are included in the sum.
{\Steve The number of input parameters being optimized is $M$. If we fit
only Al-Mg formation times, $M = 1$ and $t_{\rm SS}$ is being optimized. If we fit Hf-W formation times as well, $M=2$ and we optimize for $t_{\rm SS}$ as well as $(\hfratio)_{\rm SS}$. If we fit Mn-Cr formation times as well, then $M=4$, because we must vary both $(\mnratio)_{\rm SS}$ and $\tau_{53}$, as discussed below.}
This sum can be restricted to subsets of the achondrites, to investigate the fit of, e.g., just volcanic angrites.
The optimal values of these quantities are those that minimize $\chisq$.

In principle, $\chisq$ is a multi-dimensional function of these $M$ inputs, and must be minimized by a global search.
In practice, we find it is sufficient to treat the minimization {\Steve by first fitting the Al-Mg, Hf-W and Pb-Pb systems, and then adjusting the Mn-Cr system.}
The uncertainty in the $\mnfivethree$ half-life is large enough that it and the value of $(\mnfivethree)_{\rm SS}$ must be optimized simultaneously.
The uncertainties in the $\altwosix$ and $\hfoneeighttwo$ half-lives, in contrast, are small enough that they can be considered fixed.

Our strategy is as follows.
First we optimize $t_{\rm SS}$ by comparing the Pb-Pb ages against the Al-Mg ages (similar to the approach in Paper I).
Al-Mg ages have the lowest uncertainties and $(\alratio)_{\rm SS}$ is already defined, so this will provide the tightest constraints on $t_{\rm SS}$.
{\Steve Second, we optimize $(\hfratio)_{\rm SS}$ by comparing Hf-W against Al-Mg and Pb-Pb formation times.}
{\Steve After optimizing these two parameters, we then optimize $(\mnratio)_{\rm SS}$ and $\tau_{53}$ by comparing Mn-Cr formation times against Al-Mg, Hf-W, and Pb-Pb formation times.}
Direct comparisons between Mn-Cr and Al-Mg formation times have been made by \citet{NyquistEtal2009}, \citet{SanbornEtal2019}, \citet{TissotEtal2017}, and others, but we compare Mn-Cr against Pb-Pb because there are 10 achondrites in Table~\ref{table:one} with both Mn-Cr and Pb-Pb data, but only five that have both Mn-Cr and Al-Mg.
We then use our updated values to derive new times of formation from the data in Table~\ref{table:one}, and repeat the calculations above until the global optimization has converged; in practice, this means only a slight adjustment to quantities after the first iteration. 
After the global optimization, we will compare the Mn-Cr and Al-Mg systems, or other pairs of systems. 

\subsection{Optimal value of $t_{\rm SS}$}

We begin by finding the optimal $t_{\rm SS}$, keeping other quantities fixed.
This means finding the value $t_{\rm SS}^{*}$ for which $\partial \chisq / \partial t_{\rm SS} = 0$.
Recalling the definition of $\Delta t_{i}$ above and recognizing that $\partial \Delta t_{{\rm Pb},i} / \partial t_{\rm SS} = 1$, we find:
\begin{equation}
t_{\rm SS}^{*} = 
\frac{
\sum_{i=1}^{A} \alpha_{i} \left( 
 \frac{t_{{\rm Pb},i} +\Delta t_{26,i}}{\sigma_{\Delta t26,i}^2}
+\frac{t_{{\rm Pb},i} +\Delta t_{53,i}}{\sigma_{\Delta t53,i}^2}
+\frac{t_{{\rm Pb},i} +\Delta t_{182,i}}{\sigma_{\Delta t182,i}^2}
\right)
}
{
\sum_{i=1}^{A} \alpha_{i} \left(
 \frac{1}{\sigma_{\Delta t26,i}^2}
+\frac{1}{\sigma_{\Delta t53,i}^2}
+\frac{1}{\sigma_{\Delta t182,i}^2}
\right)
},
\label{eq:tss}
\end{equation}
where 
\begin{equation}
\alpha_i = \frac{
1 / \sigma_{\Delta t {\rm Pb},i}^2
}
{
1 / \sigma_{\Delta t {\rm Pb},i}^2
+ 1 / \sigma_{\Delta t26, i}^2
+ 1 / \sigma_{\Delta t53, i}^2
+ 1 / \sigma_{\Delta t182, i}^2
},
\end{equation}
and it is again understood that the summation is over only the relevant samples, and that for isotopic systems without data, effectively $\sigma$ is infinite. 
The optimal Pb-Pb age of $t\!\!=\!\!0$ is thus seen to be a weighted mean of the ages {\EDIT found} by adding the inferred times of formation of sample $i$ ($\Delta t_{26,i}$ or $\Delta t_{53,i}$ or $\Delta t_{182,i}$) to the Pb-Pb age of sample $i$ ($t_{{\rm Pb},i}$).
{\EDIT We derive Equation~\ref{eq:tss} in the Appendix.}

In the limit that no Mn-Cr or Hf-W data exist, only Al-Mg data, we find
\begin{equation}
t_{\rm SS}^{*} = 
\left[ \sum_{i=1}^{A} w_i \right]^{-1} \times 
\left[ 
\sum_{i=1}^{A} 
 w_i \, \left( t_{{\rm Pb},i} +\Delta t_{26,i} \right) 
 \right],
\label{eq:tsssimple}
\end{equation}
where $w_{i}^2 = 1 / \left( \sigma_{\Delta t26,i}^2 +\sigma_{\Delta t{\rm Pb},i}^2 \right)$, as found in Paper I.
Equation~\ref{eq:tss} is thus seen to be a more general form of Equation~\ref{eq:tsssimple} from Paper I, and should be used, although in practice the inclusion of Mn-Cr and Hf-W data changes $t_{\rm SS}^{*}$ only by {\EDIT $< 0.06$ Myr}.

In principle, half-lives of $\altwosix$, ${}^{235}{\rm U}$, and ${}^{238}{\rm U}$ could be found that optimize the fit.
However, the Pb-Pb ages we have included have already implicitly assumed fixed half-lives for ${}^{235}{\rm U}$ and ${}^{238}{\rm U}$, most commonly $t_{1/2} = 703.81 \pm 0.96 (1\sigma)$ Myr and $t_{1/2} = 4468.3 \pm 4.8 (1\sigma)$ Myr, respectively \citep{JaffeyEtal1971,VillaEtal2016}.
We investigate the sensitivity of the results to uncertainties in $\tau_{26}$ in \S 4.2, but we fix the $\altwosix$ half-life at 0.717 Myr during the optimization procedure.

\subsection{Optimal value of $(\hfratio)_{\rm SS}$}

We next find the optimal value of $R_{182,{\rm SS}} = (\hfratio)_{\rm SS}$ that minimizes $\chisq$.
That is, we find $R_{182}^{*}$ for which $\partial \chisq / \partial R_{182,{\rm SS}} = 0$.
Direct differentiation of $\chisq$ yields a cumbersome result unless it is assumed that $\Delta t_{i}$ is insensitive to $R_{182,{\rm SS}}$.
Fortunately, this is the case, as in almost all instances, $\Delta t_{i}$ is much more heavily weighted to $\Delta t_{26,i}$ and $\Delta t_{{\rm Pb},i}$ than to $\Delta t_{182,i}$, the latter having the largest age uncertainty due to the long half-life of $\hfoneeighttwo$.
Accordingly, we find: 
\begin{equation}
\!\!\!\!\!\!\!\!\!\!\!\!\!\!\!\!\!\!\!\!\!\!\!
\!\!\!\!\!\!\!\!\!\!\!\!\!\!\!\!\!\!\!\!\!\!\!
\!\!\!\!\!\!\!\!\!\!\!\!\!\!\!\!\!\!\!\!\!\!\!
\!\!\!\!\!\!\!\!\!\!\!\!\!\!\!\!\!\!\!\!\!\!\!
\!\!\!\!\!\!\!\!\!\!\!\!\!\!\!\!\!\!\!\!\!\!\!
\!\!\!\!\!\!\!\!\!\!\!\!\!\!\!\!\!\!\!\!\!\!\!
(\hfratio)_{\rm SS}^{*} = 
\end{equation}
\[
\exp \left\{ \left[ \sum_{i=1}^{A} w_{i} \right]^{-1} \times \left[ \sum_{i=1}^{A} \, w_i \, \left( \ln (\hfratio)_{0,i} + \frac{ \Delta t_{i} }{ \tau_{182} } \right) \right] \right\}, 
\]
where $w_i = 1 / \sigma_{\Delta t182,i}^2$.

The measured half-life of $\hfoneeighttwo$ is $8.896 \pm 0.089 (1\sigma)$ Myr \citep{VockenhuberEtal2004}, so the $2\sigma$ uncertainty in the half-life is only 2\%. 
In \S 4.2 we investigate the sensitivity of the results to $\tau_{182}$, but during the optimization we fix the half-life at 8.896 Myr.

\subsection{Optimal values of $(\mnratio)_{\rm SS}$ and $\tau_{53}$}

We next find the optimal values of $R_{53,{\rm SS}}$ and $\tau_{53}$ that minimize $\chisq$. 
That is, we find $R_{53,{\rm SS}}^{*}$ and $\tau_{53}^{*}$ for which $\partial \chisq / \partial R_{53,{\rm SS}} = 0$ and $\partial \chisq / \partial \tau_{53} = 0$ simultaneously.

We again take advantage of the fact that $\Delta t_{i}$ is much more tied to Al-Mg formation times $\Delta t_{26,i}$, or the Pb-Pb formation times $\Delta t_{{\rm Pb},i}$ that are tied to them, than to Mn-Cr formation times.
In that case, assuming fixed $\tau_{53}$,
\begin{equation}
\!\!\!\!\!\!\!\!\!\!\!\!\!\!\!\!\!\!\!\!\!\!\!
\!\!\!\!\!\!\!\!\!\!\!\!\!\!\!\!\!\!\!\!\!\!\!
\!\!\!\!\!\!\!\!\!\!\!\!\!\!\!\!\!\!\!\!\!\!\!
\!\!\!\!\!\!\!\!\!\!\!\!\!\!\!\!\!\!\!\!\!\!\!
\!\!\!\!\!\!\!\!\!\!\!\!\!\!\!\!\!\!\!\!\!\!\!
\!\!\!\!\!\!\!\!\!\!\!\!\!\!\!\!\!\!\!\!\!\!\!
(\mnratio)_{\rm SS}^{*} = 
\end{equation}
\[
\exp \left\{ \left[ \sum_{i=1}^{A} w_{i} \right]^{-1} \times \left[ \sum_{i=1}^{A} \, w_i \, \left( \ln (\mnratio)_{0,i} + \frac{ \Delta t_{i} }{ \tau_{53} } \right) \right] \right\}, 
\]
where $w_i = 1 / \sigma_{\Delta t53,i}^2$.

Conversely, assuming fixed $(\mnratio)_{\rm SS}$, we find
\begin{equation}
\tau_{53}^{*} = \frac{
\sum_{i=1}^{A} w_i \, \Delta t_{i} \, \ln \left( R_{53,{\rm SS}} / R_{53,i} \right) }
{ \sum_{i=1}^{A} w_i \, \left[ \ln \left( R_{53,{\rm SS}} / R_{53,i} \right) \right]^2 },
\end{equation}
where again $w_i = 1 / \sigma_{\Delta t53,i}^2$.

Because of the coupled nature of the problem, we first assume a value for $\tau_{53}$ and find $R_{53,{\rm SS}}^{*}$; then, setting $R_{53,{\rm SS}} = R_{53,{\rm SS}}^{*}$, we find the optimal value $\tau_{53}^{*}$; then, refining $\tau_{53}$ to be equal to this new $\tau_{53}^{*}$, we again find $R_{53,{\rm SS}}^{*}$ and iterate this procedure to convergence.

{\Steve 
\subsection{Searching for an optimal fit}

The equations above for $t_{\rm SS}^{*}$, $(\hfratio)_{\rm SS}^{*}$, $(\mnratio)_{\rm SS}^{*}$ and $\tau_{53}^{*}$ yield a fast way to search the multi-variable parameter space. 
In practice, however, we also can vary the four input parameters across a four-dimensional grid of input parameters and find the minimum $\chisq$ by brute force.
Testing $\sim 10^8$ combinations of input parameters takes less than one minute on a typical laptop computer. 
We find the two approaches yield almost identical answers, with differences in $\chisq$ at the $< 1\%$ level.
}

\subsection{Statistical significance of a fit}

{\Steve 
As in Paper I, the significance of a fit is first assessed by considering the $z$ scores of various ages, e.g., $z_{26,i} = 2(\Delta t_{26,i} - \Delta t_{i}) / \sigma_{\Delta t26}$.
In an acceptable fit, most (95\%) but not all $z$ scores should be characterized by $\left| z \right| < 2$.}

{\Steve The fit is also assessed in a global sense through $\chisq$.}
The procedure above finds the minimum value of $\chisq$ given fixed half-lives of $\altwosix$ and $\hfoneeighttwo$, allowing variations in four parameters: $t_{\rm SS}$, $(\mnratio)_{\rm SS}$, $\tau_{53}$, and $(\hfratio)_{\rm SS}$.
{\Steve Here $\nu$ is the number of degrees of freedom, equal to $N-4$ if we have $N$ times of formation being fit simultaneously, using 4 free parameters.}
If they do match, then we should see $\chisq \approx 1$.
{\Steve In general, the probability that the data are concordant but have $\chisq > 1$ as high as they do due to measurement errors alone is the cumulative distribution $P_{\nu}(>\chisq)$  \citep{WendtCarl1991}.
Closed-form solutions exist. 
The usual (``$2\sigma$") threshold for acceptance is that this probability exceed 5\%, which places an upper limit to $\chisq$ such that $P_{\nu}(\chisq_{\rm max}) = 0.05$.
For example, if $\nu=16$, then $\chisq_{\rm max} = 1.644$.}
As discussed by \citet{WendtCarl1991}, in the limit of large $N$, if the data scatter is only due to (Gaussian) measurement error, then the probability is 95\% that $\chisq_{\rm min} < \chisq < \chisq_{\rm max}$, where 
\begin{equation}
\chisq_{\rm max} \approx 1 + 2 \, \left( \frac{2}{N-M} \right)^{1/2},
\label{eq:chisqmax}
\end{equation}
where again $N-M$ reflects the fact that there are $N$ data fit with $M$ input parameters, so there are $N-M$ degrees of freedom. 
There is only a $< 5\%$ probability that $\chisq > \chisq_{\rm max}$ (or less than a corresponding function $\chisq_{\rm min}$).
{\Steve Equation~\ref{eq:chisqmax} is then approximate, as $\chisq_{\rm max} = 1.707$ for $\nu = 16$, compared to 1.644. }

If some subset of the achondrites above yields $\chisq < \chisq_{\rm max}$, we consider this a satisfactory fit, and the isotopic systems can be considered concordant in those achondrites. 
Such a fit would fail to invalidate the two assumptions of homogeneity of $\altwosix$, $\mnfivethree$ and $\hfoneeighttwo$, and that the isotopic systems achieved closure in each sample. 
If some subset of achondrites yields $\chisq > \chisq_{\rm max}$, then {\it either} the assumption of homogeneity has been violated, {\it or} the closures of some isotopic systems were not simultaneous in one or more of the achondrites, or both.
Petrologic context would be necessary to assess these possibilities.

\section{Statistical Chronometry Results}

\subsection{Volcanic achondrites}

Because the assumption of simultaneously closed isotopic systems is more likely to be satisfied in the volcanic achondrites than other achondrites, we begin our analysis with the three quenched angrites D'Orbigny, SAH 99555, and NWA 1670, the eucrite-like Asuka 881394, and NWA 7325,
{\Steve and the rapidly cooled CC achondrites NWA 2976 and NWA 6704.
These are the same seven achondrites considered in Paper I, for which the Al-Mg and Pb-Pb formation times were shown to be concordant.
We now ask whether they remain concordant when considering their Mn-Cr and Hf-W ages as well.
}

{\Steve We first consider only the Al-Mg and Pb-Pb systems.
Across these seven achondrites, there are 14 formation times $\Delta t_{26}$ and $\Delta t_{\rm Pb}$ which we fit using the single parameter $t_{\rm SS}$.
We find an optimal fit $t_{\rm SS} = 4568.377$ Myr, as in Paper I, with $\chisq = 0.979$, which is an excellent fit, with $P(\chisq) = 47\%$.
}

{\Steve 
We next consider the effects of including Hf-W ages, which exist for D'Orbigny and SAH 99555.
Across these seven achondrites there are now 16 formation times to be fit using only the two parameters $t_{\rm SS}$ and $(\hfratio)_{\rm SS}$.
We find an optimal fit 
\begin{eqnarray}
(\hfratio)_{\rm SS}^{*} & = & 10.402 \times 10^{-5} \nonumber \\
t_{\rm SS}^{*} & = & 4568.370 \, {\rm Myr}. \nonumber
\end{eqnarray}
with $\chisq = 1.24$, also a good fit, with $P(\chisq) = 24\%$. 
These seven achondrites easily can be considered concordant across their Al-Mg, Hf-W, and Pb-Pb ages.

An examination of the $z$ scores of the formation times strengthens the case for concordance. 
The Pb-Pb age of SAH 99555 is discordant at the $2.2\sigma$ level, and the Pb-Pb age of NWA 6704 at almost the $2.0\sigma$ level; but this is exactly as expected.
Assuming these ages are scattered only due to their reported measurement uncertainties, we would expect 68\% of the 16 ages, i.e., 10.9, to have $\left| z \right| < 1$; 27\%, or 4.3, to have $1 < \left| z \right| < 2$, 5\%, or 0.8, to have $2 < \left| z \right| < 3$, and $0.003\%$, or 0.05, to have $\left| z \right| > 3$.
The actual distributions are 12, 3, 1, and 0.
}

{\Steve 
We now consider the effects of including the Mn-Cr formation times that exist for D'Orbigny, SAH 99555, NWA 1670, and Asuka 881394.
Across these seven} achondrites there are $N\!\!=\!\!21$ formation times that must be fit simultaneously.
demanding $\chisq < \chisq_{\rm max} \approx 1.69$.

Fitting the four parameters simultaneously, we find the following global minimum in $\chisq$:
\begin{eqnarray}
\tau_{53}^{*} & = & 6.70 \, {\rm Myr} \nonumber \\
\, & \, & (t_{1/2} = 4.64 \, {\rm Myr}) \nonumber \\
(\mnratio)_{\rm SS}^{*}  & = & 6.87 \times 10^{-6} \nonumber \\
(\hfratio)_{\rm SS}^{*} & = & 10.40 \times 10^{-5} \nonumber \\
t_{\rm SS}^{*} & = & 4568.37 \, {\rm Myr}.\nonumber
\end{eqnarray}
For these values, $\chisq = 1.59$, which is an acceptable fit
{\Steve (6\% probability). We do not assign significance to this solution, though, because of the aphysical $\mnfivethree$ half-life.
In fact, a half-life of 3.8 Myr would fit the data almost equally well, with $\chisq = 1.69$ and $P(\chisq) = 5\%$.
We defer discussion of Mn-Cr systematics until after considering other achondrites.
}

{\Steve 
\subsection{All achondrites}

We now consider whether the Al-Mg, Hf-W, and Pb-Pb formation times will remain concordant if we include other achondrites into the statistical test. 
We now add the six plutonic angrites LEW 86010, NWA 1296, NWA 2999, NWA 4590, NWA 4801, and Angra dos Reis (AdoR), to the list of seven achondrites above. 
Among these 12 achondrites there are 26 formation times to be fit using only the same two parameters, $t_{\rm SS}$ and $(\hfratio)_{\rm SS}$, as above. 
An acceptable fit requires an even smaller $\chisq$, less than about 1.58.

Amazingly, we find an acceptable optimal solution:
\begin{eqnarray}
(\hfratio)_{\rm SS}^{*} & = & 10.500 \times 10^{-5} \nonumber \\
t_{\rm SS}^{*} & = & 4568.326 \, {\rm Myr}. \nonumber
\end{eqnarray}
For these values, $\chisq = 1.31$, which is a good fit  (14\% probability).
The $z$ scores also appear distributed normally, except for NWA 4801; its Hf-W formation time is discordant at the $2.9\sigma$ level. 

If we don't include the Hf-W formation time of NWA 4801, we can fit 11 achondrites with 24 formation times using only the same two parameters, $t_{\rm SS}$ and $(\hfratio)_{\rm SS}$, as above. 
An acceptable fit requires an even smaller $\chisq < 1.60$.
Now the optimal fit is 
\begin{eqnarray}
(\hfratio)_{\rm SS}^{*} & = & 10.427 \times 10^{-5} \nonumber \\
t_{\rm SS}^{*} & = & 4568.360 \, {\rm Myr}. \nonumber
\end{eqnarray}
For these values, $\chisq = 0.959$, which is an exceptionally good fit, with $P(\chisq) = 51.4\%$.
The $z$ scores are also distributed normally: 19 with $\left| z \right| < 1$, 4 with $\left| z \right| = 1 - 2$, 1 with $\left| z \right| = 2-3$, and 0 with $\left| z \right| > 3$.
With the exception of the Hf-W formation time of NWA 4801, the Al-Mg, Hf-W, and Pb-Pb systems appear concordant across all 11 achondrites, including the CC achondrites and the quenched and plutonic angrites.

We can vary these two parameters about these optimal values and see which combinations yield solutions with $> 5\%$ probability, which demands $\chisq < \chisq_{\rm max} \approx 1.58$ (actually, 1.541). 
For fixed $(\hfratio)_{\rm SS}$, we find the acceptable range is $t_{\rm SS} \approx 4568.36 \pm 0.24$ Myr.
For fixed $t_{\rm SS}$, we find the acceptable range of $(\hfratio)_{\rm SS}$ is $\approx (10.42 \pm 0.28) \times 10^{-5}$.
Across these ranges, the Al-Mg, Hf-W and Pb-Pb systems remain concordant with each other. 
}

{\Steve
\subsection{Mn-Cr ages}
}

{\Steve
We now consider the concordancy of the Mn-Cr formation times across all the achondrites, rejecting only the Hf-W formation time of NWA 4801.
We include all Mn-Cr formation times, yielding 37 formation times across 14 achondrites.
We seek values of 
 $t_{\rm SS}$, $(\hfratio)_{\rm SS}$, $(\mnratio)_{\rm SS}$, and the $\mnfivethree$ half-life that optimize the fit.
The range of values of $(\hfratio)_{\rm SS}$ and $t_{\rm SS}$ that provide an acceptable fit are quite restricted and we fix them at the  values above.
In contrast, there are many combinations of $(\mnratio)_{\rm SS}$ and $\mnfivethree$ half-life that provide acceptable fits.
We have calculated $\chisq$ across a grid of values for the $\mnfivethree$ half-life and the initial $(\mnratio)_{\rm SS}$ ratio in the Solar System.
For each value of $\chisq$ we calculate the probability $P_{\rm fit}(\chisq)$ that the fit would have that $\chisq$, using the formulas of \citep{WendtCarl1991} and assuming 33 degrees of freedom.
In Figure~\ref{fig:trough}a we plot $\log_{10}[P_{\rm fit}(\chisq)]$. 
Values $> -1.30$ are statistically significant at the $2\sigma$ level ($> 5\%$
probability). 
The minimum $\chisq$ surface in this two-dimensional space of $(\mnratio)_{\rm SS}$ vs.\ $\tau_{53}$ describes a long and narrow trough.
Most of the constraints on the Mn-Cr system come from quenched angrites like D'Orbigny that formed at around $\Delta t \approx 5$ Myr, so empirically $(\mnratio)_{\rm SS} \, \exp( - (5 \, {\rm Myr} / \tau_{53} ) \approx 3.24 \times 10^{-6}$.
For a given half-life, the uncertainty in $(\mnratio)_{\rm SS}$ is typically about $\pm(0.20) \times 10^{-6}$.
It is apparent that half-lives anywhere from $< 3$ to $> 5$ Myr could fit the data.
This is aphysical, though, as experiments constrain the half-life to a range 3.1 to 4.3 Myr. 
}

{\Steve
We next calculate the range of half-lives that are consistent with both the chronometry {\it and} the experimental determinations. 
To do so we calculate a new probability $P_{1/2} = 0.3989 \, \exp \left[ -(t_{1/2} -3.70 \, {\rm Myr})^2 / 2 (0.31 \, {\rm Myr})^2 \right]$, due to experimental determination of the half-life.
In Figure~\ref{fig:trough}b we plot the joint probability $P_{\rm fit} \times P_{1/2}$ as a function of $(\mnratio)_{\rm SS}$ and the $\mnfivethree$ half-life. 
The most probable value of the $\mnfivethree$ half-life is about 3.80 Myr and the probable ($> 5\%$) range is restricted to {\EDIT 3.57} to 4.14 Myr, or roughly $3.80 \pm 0.23(2\sigma)$ Myr.
The most likely value 3.80 is not very different from the commonly adopted 3.7 Myr, but has half the uncertainty.
Half-lives from 3.1 to 3.5 Myr are allowed by the experiments, but are less probable; in combination with the improbability of matching the chronometry, they can be ruled out. 
Likewise, half-lives from 4.1 to $> 5$ Myr make the formation times concordant, but are not also consistent with the experimental half-life.
Under the assumption of uniform $\mnfivethree$, the range of half-lives consistent with both the chronometry and the experiments is {\EDIT about 3.6} to about 4.1 Myr.
It could have been the case that the only half-lives making the Mn-Cr ages consistent was outside the range allowed by experiments; this would have validated the assumption of homogeneous $\mnfivethree$, but instead this assumption is supported.
}

{\Steve
With the range of likely half-lives constrained, the range of $(\mnratio)_{\rm SS}$ values consistent with concordant ages is also found. 
The allowed range of $(\mnratio)_{\rm SS}$ is $(8.09_{-0.59}^{+0.71}) \times 10^{-6}$.
It is understood that $\mnfivethree$ and $(\mnratio)_{\rm SS}$ cannot be varied independently and still fit the formation times.
}

{\Steve
If we adopt a half-life 3.80 Myr, we find the following optimal fit:
\begin{eqnarray}
\tau_{53}^{*} & = & 5.482 \, {\rm Myr} \nonumber \\
\, & \, & (t_{1/2} = 3.80 \, {\rm Myr}) \nonumber \\
(\mnratio)_{\rm SS}^{*} & = & 8.09 \times 10^{-6} \nonumber \\
(\hfratio)_{\rm SS}^{*} & = & 10.421 \times 10^{-5} \nonumber \\
t_{\rm SS}^{*} & = & 4568.355 \, {\rm Myr}. \nonumber
\end{eqnarray}
The fit is completely concordant.
For this combination, $\chisq = 1.09$, indicating an excellent fit (33\% probability). 
The $z$ scores appear to be distributed normally. 
Among 37 formation times, we would expect 25.2 to be discordant by $< 1 \sigma$, 10.0 to be discordant by 1 to $2\sigma$, 1.7 to be discordant by 2 to $3 \sigma$, and 0.1 to be discordant by $> 3\sigma$. 
We find 24, 11, 2, and 0. 
The two most discordant ages are the Pb-Pb age of SAH 99555 (discrepant at the $2.3\sigma$ level) and the Mn-Cr age of NWA 6704 (discrepant at the $2.4\sigma$ level).
In summary, all 37 robust formation times (except the Hf-W formation time of NWA 4801) are concordant.
}

\begin{center}
\begin{figure}[ht!]
\includegraphics[width=0.49\textwidth,angle=0]{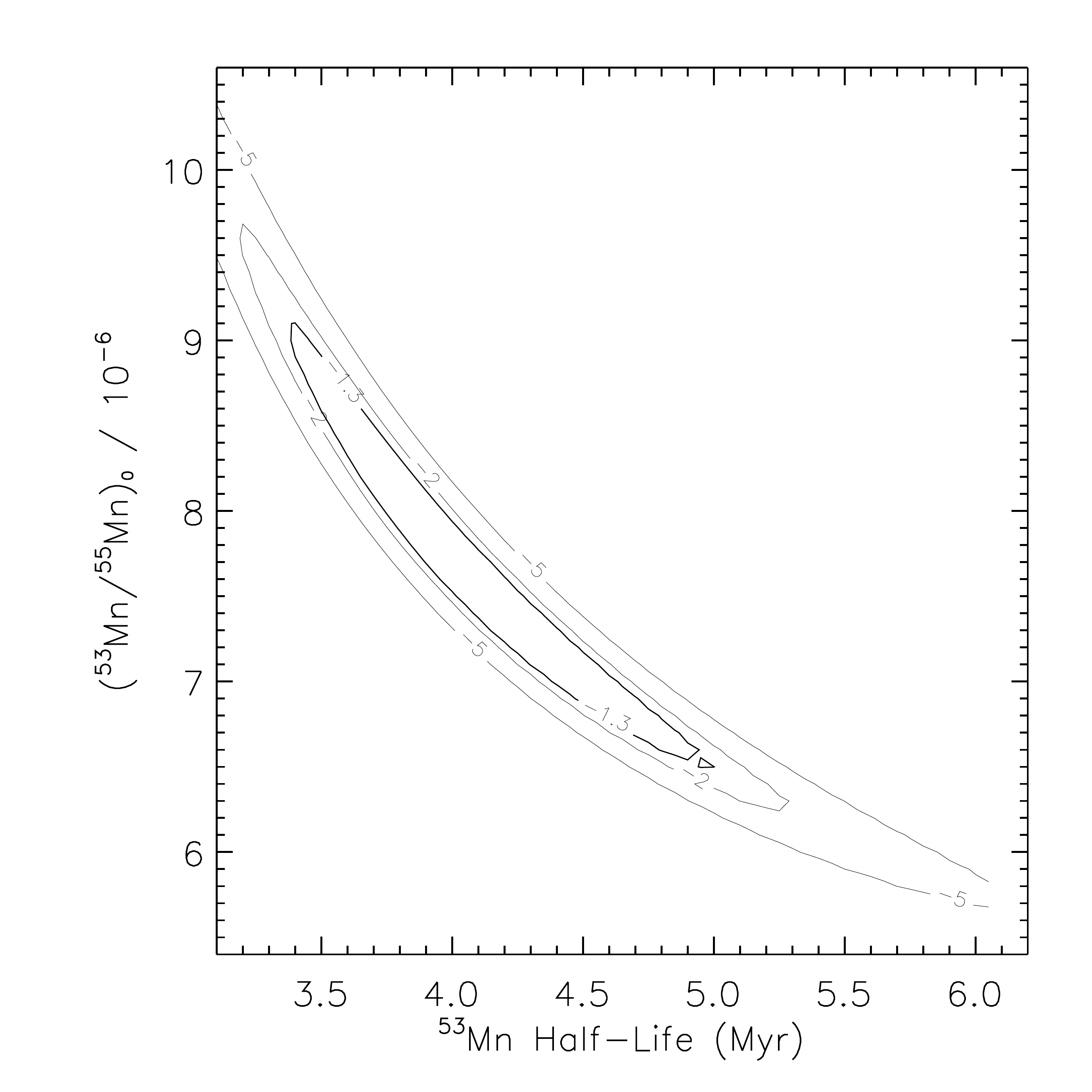}
\includegraphics[width=0.49\textwidth,angle=0]{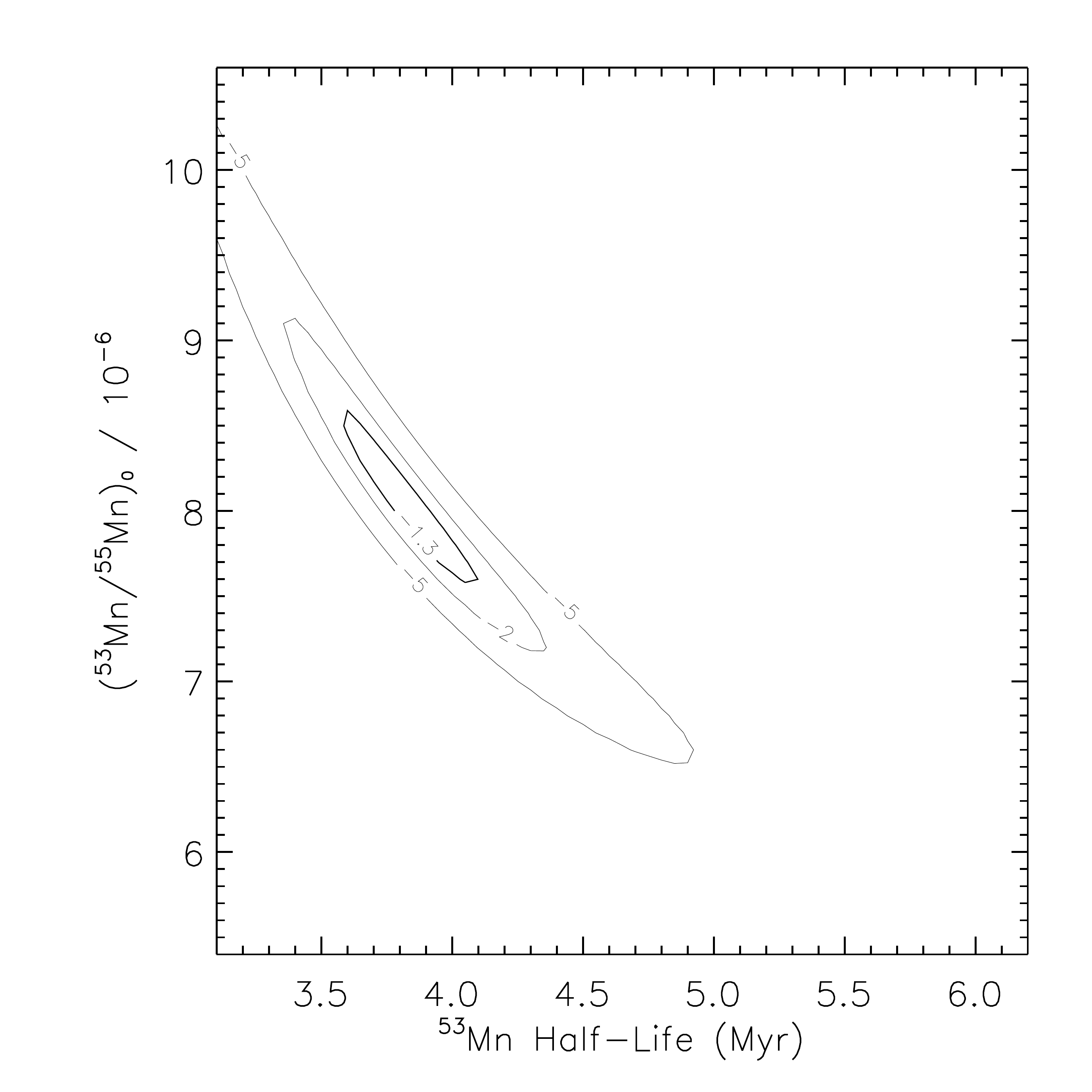}
\caption{{\it Left.} Logarithm of the probability of the fit as a function of the input parameters, the assumed half-life of $\mnfivethree$ and the initial ratio $(\mnratio)_0$. Values $> -1.3$ (bold contour) are statistically significant ($> 5\%$ probability).
{\it Right.} The joint probability including the experimental constraints on the half-life. 
\label{fig:trough}}
\end{figure}
\end{center}

{\Steve
Fixing the parameters above, we can investigate the discrepancies of the ages we have excluded.
{\EDIT For NWA 2999, based on the reported} $(\mnratio)_{0} = (1.28 \pm 0.23) \times 10^{-6}$ \citep{ShukolyukovLugmair2008}, we would find $\Delta t_{53} = 10.11 \pm 0.99$ Myr, which differs from our computed $\Delta t = 7.95$ Myr by $> 4\sigma$.
This value was not published in the refereed literature.
\citet{LugmairShukolyukov1998} had regressed two Mn-Cr data points for AdoR along with data for LEW 86010 and reported $(\mnratio)_0 = (1.25 \pm 0.07) \times 10^{-6}$. 
This value and uncertainty would yield the same Mn-Cr formation time $\Delta t_{53} = 9.87 \pm 0.20$ {\EDIT Myr for AdoR} as for LEW 86010, which is discrepant from AdoR's $\Delta t = 12.08$ Myr by $> 22\sigma$.
Based on Mn-Cr and Hf-W data, we calculate a time of formation of LEW 86010 of $\Delta t = 9.88$ Myr.
Its Pb-Pb age is $4558.55 \pm 0.15$ Myr, based on a U isotopic ratio 137.88 \citep{Amelin2008b}, yielding $\Delta t_{\rm Pb} = 9.80 \pm 0.15$.
To be concordant to within $2\sigma$, it should be younger than 4558.55 Myr, by about 0.07 Myr, but not more than 0.22 Myr.
This is comparable to the age correction found using the ${}^{238}{\rm U}/{}^{235}{\rm U}$ ratio of the similar plutonic angrite NWA 4590. 
These data previously flagged as questionable indeed turn out not to fit with the other data.
}

{\Steve
This leaves only the case of the Hf-W formation time of NWA 4801.
Including it in the optimization slightly changes the inferred parameters [e.g., $\mnfivethree$ half-life $\equiv 3.80$ Myr, $(\mnratio)_{\rm SS} = 8.093 \times 10^{-6}$, $(\hfratio)_{\rm SS} = 10.496 \times 10^{-5}$, $t_{\rm SS} = 4568.320$ Myr], but the difference makes the fit for SAH 99555 worse, with $\Delta t_{182}$ and $\Delta t_{\rm Pb}$ discrepant at the $2.2\sigma$ and $2.6\sigma$ levels, and now $\Delta t_{53}$ for NWA 6704 is discrepant by $2.5\sigma$.
Although the ages of NWA 4801 are not terribly discordant [its inferred formation time $\Delta t_{182} = 10.81$ Myr is discrepant from its inferred $\Delta t = 11.47$ Myr by $2.9\sigma$],
including NWA 4801 in the fit makes the overall fit much worse, with $\chisq = 1.408$.
While below the threshold $1.50$ for statistical significance, the probability of the fit is only 6\%, much worse than the probability of 33\% found without including NWA 4801. 
\citet{McKibbinEtal2015} considered this achondrite disturbed, noting the description of it by \citet{IrvingKuehner2007} as an annealed breccia formed originally by disruption of a very coarse-grained plutonic protolith, implying a late thermal annealing event that could have reset some isotopic systems.
\citet{KleineEtal2012} also noted it had an unusually high abundance of W in its fines fraction. 
We choose to exclude NWA 4801 from our optimization when calculating Solar System values, but note that technically the chronometry would be concordant even including it.
}


\subsection{Sensitivity to Parameters}

{\Steve
Around this optimal fit, we can examine what ranges of input parameters yield acceptable fits.
We first keep variables fixed around the values above and vary them one at a time, finding values that make the 37 ages concordant ($\chisq < 1.44$). 
We find:
\begin{eqnarray}
\tau_{53}^{*} & = & 5.48 \, \pm 0.33 \, {\rm Myr} \nonumber \\
\, & \, & (t_{1/2} = 3.80 \pm 0.23 \, {\rm Myr}) \nonumber \\
(\mnratio)_{\rm SS}^{*} & = & (8.09_{-0.58}^{+0.71}) \times 10^{-6} \nonumber \\
(\hfratio)_{\rm SS}^{*} & = & (10.42 \pm 0.23) \times 10^{-5} \nonumber \\
t_{\rm SS}^{*} & = & 4568.35 \pm 0.19 \, {\rm Myr}. \nonumber
\end{eqnarray}
}

In our analysis, we formally allowed only four parameters to vary: $t_{\rm SS}$, $(\mnratio)_{\rm SS}$, $\tau_{53}$, and $(\hfratio)_{\rm SS}$.
We fixed $\tau_{26} = 1.0344$ Myr ($t_{1/2} = 0.717$ Myr) and $\tau_{182} = 12.834$ Myr ($t_{1/2} = 8.90$ Myr), but equations similar to Equation 22 can be used to query what the optimal values of $\tau_{26}$ and $\tau_{182}$ should be.
Varying the half-life of $\altwosix$ (and optimizing the other variables), we find that the optimal fit is for values near 0.717 Myr, but that no meaningful variations in $\chisq$ occur over the experimentally allowed range.
The optimal value of $t_{\rm SS}$ does change, however, with variations in $\tau_{26}$ across its $\pm 0.017$ Myr range yielding variations in $t_{\rm SS}$ of $\pm 0.14$ Myr.
Varying the half-life of $\hfoneeighttwo$ across its experimentally allowed range does not change the fit almost at all, with variations in $t_{\rm SS}$ less than $\pm 0.01$ Myr.


{\Steve
Finally, we examine the effects of the age correction of 0.19 Myr applied to all achondrites {\it other} than the CC achondrites (NWA 2976 and NWA 6704).
As discussed in Paper I, we take the fraction of U in pyroxenes, $f_{\rm cpx}$, to be 1 in those achondrites and, following \citet{TissotEtal2017}, 0.5 in angrites and other NC achondrites. 
If we do not apply this correction to the NC achondrites (setting $f_{\rm cpx} = 1$), their ages would be 0.19 Myr older than our optimal solution, and $t_{\rm SS}$ would increase by about 0.14 Myr, but worsening the fit: $\chisq$ would increase from 1.09 to 1.32, and the Pb-Pb age of SAH 99555 made discordant at the $2.7\sigma$ level. 
If we apply the maximum correction of 0.38 Myr ($f_{\rm cpx} = 0$) to the NC achondrites, they would be 0.19 Myr younger than our optimal solution, and $t_{\rm SS}$ would decrease by about 0.14 Myr, and overall the fit would be much improved, with $\chisq = 0.96$, although now the Pb-Pb age of D'Orbigny would be discordant at the $2.6\sigma$ level.
These findings suggest that the correction for the isotopic fractionation between pyroxenes and whole-rock U isotopic compositions advocated by \citep{TissotEtal2017} is necessary, i.e., current ages of NC achondrites are overestimated, by at least 0.2 Myr. 
}



\subsection{Summary}

{\Steve
We have compiled and vetted nearly 40 reported formation times measured by four isotopic systems, across 14 achondrites.
We find them remarkably concordant.
Only the Hf-W formation time of NWA 4801 appears not to fit, possibly because it suffered a late-stage annealing event \citep{IrvingKuehner2007}, possibly associated with the unusually high W abundance in its matrix \citep{KleineEtal2012}.
Two previously repoted Mn-Cr ages of plutonic angrites could not be reconciled, but the one for NWA 2999 was not in the refereed literature, and the one for Angra dos Reis was actually conflated with the Mn-Cr isochron for LEW 86010.
The Pb-Pb age of LEW 86010 was not U-corrected, but would be reconciled with the other formation times if its U isotopic ratio were the same as NWA 4590, another similar plutonic angrite.
The other 37 formation times across 14 achondrites are made remarkably concordant by optimizing only four parameters.
Across the following ranges, the solution is concordant:

\begin{eqnarray}
\mbox{\boldmath$t_{1/2,53}$} & = & \mbox{\boldmath$3.80 \pm 0.23 \, {\rm Myr}$} \nonumber \\
\mbox{\boldmath$(\mnratio)_{\rm SS}^{*}$} & = & \mbox{\boldmath$(8.09 \pm 0.65) \times 10^{-6}$} \nonumber \\
\mbox{\boldmath$(\hfratio)_{\rm SS}^{*}$} & = & \mbox{\boldmath$(10.42 \pm 0.23) \times 10^{-5}$} \nonumber \\
\mbox{\boldmath$t_{\rm SS}^{*}$} & = & \mbox{\boldmath$4568.35 \pm 0.19 \, {\rm Myr}$}. \nonumber
\end{eqnarray}
Uncertainties are $2\sigma$.
The values of $(\mnratio)_{\rm SS}$ and the $\mnfivethree$ half-life are not to be varied independently; if the half-life were known, then the uncertainty in $(\mnratio)_{\rm SS}$ would be only $\pm(0.20) \times 10^{-6}$.
The center of these ranges is a concordant solution, with $\chisq = 1.16$ (42\% probability solution) and deviations distributed normally. 
}

Using these values, we calculate the following times of formation for all the achondrites, listed in {\bf Table~\ref{table:two}} and depicted in {\bf Figure~\ref{fig:achondrites}}.
{\Steve For NWA 4801, we calculated a weighted mean $\Delta t$ using only the Mn-Cr and Pb-Pb ages.}

\begin{table}[htbp!]
\noindent
\begin{minipage}{5.9in}
\footnotesize
\caption{Inferred times of formation after $t\!\!=\!\!0$, in Myr, of 14 achondrites according to the Al-Mg, Mn-Cr, Hf-W, and Pb-Pb systems, and weighted means of formation times, using the parameters that optimize the fit to all but NWA 4801.  Italics denote formation times differing by $> 2\sigma$ from the inferred formation time of the achondrite.\label{table:two}}
\bigskip
\begin{tabular}{l|rr|rr|rr|rr|rr}

\hline \hline \\
{\bf Achondrite} & $\Delta t_{26}$ & $2\sigma$ & 
$\Delta t_{53}$  & $2\sigma$ &
$\Delta t_{182}$ & $2\sigma$ &
$\Delta t_{\rm Pb}$ & $2\sigma$ &
$\Delta t$ & $2\sigma$ \\ \hline
{\bf D'Orbigny} & 5.06 & 0.10 & 5.03 & 0.06 & 4.84 & 0.31 & 5.12 & 0.21 & {\bf 5.03} & {\bf 0.05} \\
{\bf SAH 99555} & 5.14 & 0.05 & 4.95 & 0.28 & 5.35 & 0.28 & {\it 4.85} & {\it 0.24} & {\bf 5.12} & {\bf 0.05} \\
{\bf NWA 1670}  & 4.64 & 0.10 & 5.72 & 1.77 & $\,$    & $\,$     & 4.34 & 0.66 & {\bf 4.63} & {\bf 0.10} \\
{\bf NWA 1296}  &  &  &  &  & 5.09    & 0.51     & 5.34 & 0.45 & {\bf 5.23} & {\bf 0.34} \\
\hline
{\bf NWA 2999}  & $\,$     &  $\,$    & {}  & {} & 8.37  & 0.80 & {7.81}  & 0.47 & {\bf 7.95}  & {\bf 0.41} \\
{\bf LEW 86010} &  $\,$    &  $\,$    & 9.84 & 0.20 & 9.95 & 1.12 &  &  & {\bf 9.84} & {\bf 0.20} \\
{\bf NWA 4590}  & $\,$     &  $\,$    & {12.35} & {2.58} & 10.41 & 0.47 & 10.79 & 0.38 & {\bf 10.66} & {\bf 0.29} \\
{\bf NWA 4801}  & $\,$     &  $\,$    & {11.93} & {1.76} & {\it 10.72} & {\it 0.45} & {11.64} & {0.21} & {\bf 11.64} & {\bf 0.21} \\
{\bf Angra dos Reis} & $\,$ & $\,$     & {10.94} & {1.99} & {12.23} & {0.77} & {12.10} & {0.29} & {\bf 12.09} & {\bf 0.27} \\
\hline
{\bf Ibitira}      &  $\,$    &  $\,$     & 11.14     &  2.59     & $\,$  &  $\,$  & 11.80 & 0.57 & {\bf 11.77} & {\bf 0.56} \\
{\bf Asuka 881394} & 3.82 & 0.04 & {4.05} & {0.32} & $\,$  & $\,$  &  3.60 & 0.53 & {\bf 3.82} & {\bf 0.04} \\
{\bf NWA 7325} & 5.33 & 0.05 & $\,$ & $\,$ & $\,$ & $\,$ & 4.65 & 1.7 & {\bf 5.33} & {\bf 0.05} \\
{\bf NWA 2976} & 5.03 & 0.04 & $\,$ & $\,$ & $\,$ & $\,$ & {5.20} & {0.57} & {\bf 5.03} & {\bf 0.04} \\
{\bf NWA 6704} & {5.29} & {0.13} & {\it 6.24} &  {\it 0.72} & $\,$ & $\,$ & {5.60} & {0.26} & {\bf 5.37} & {\bf 0.11} \\ \hline
\end{tabular}
\bigskip
\end{minipage}
\end{table}

\begin{center}
\begin{figure}[ht!]
\includegraphics[width=0.99\textwidth,angle=0]{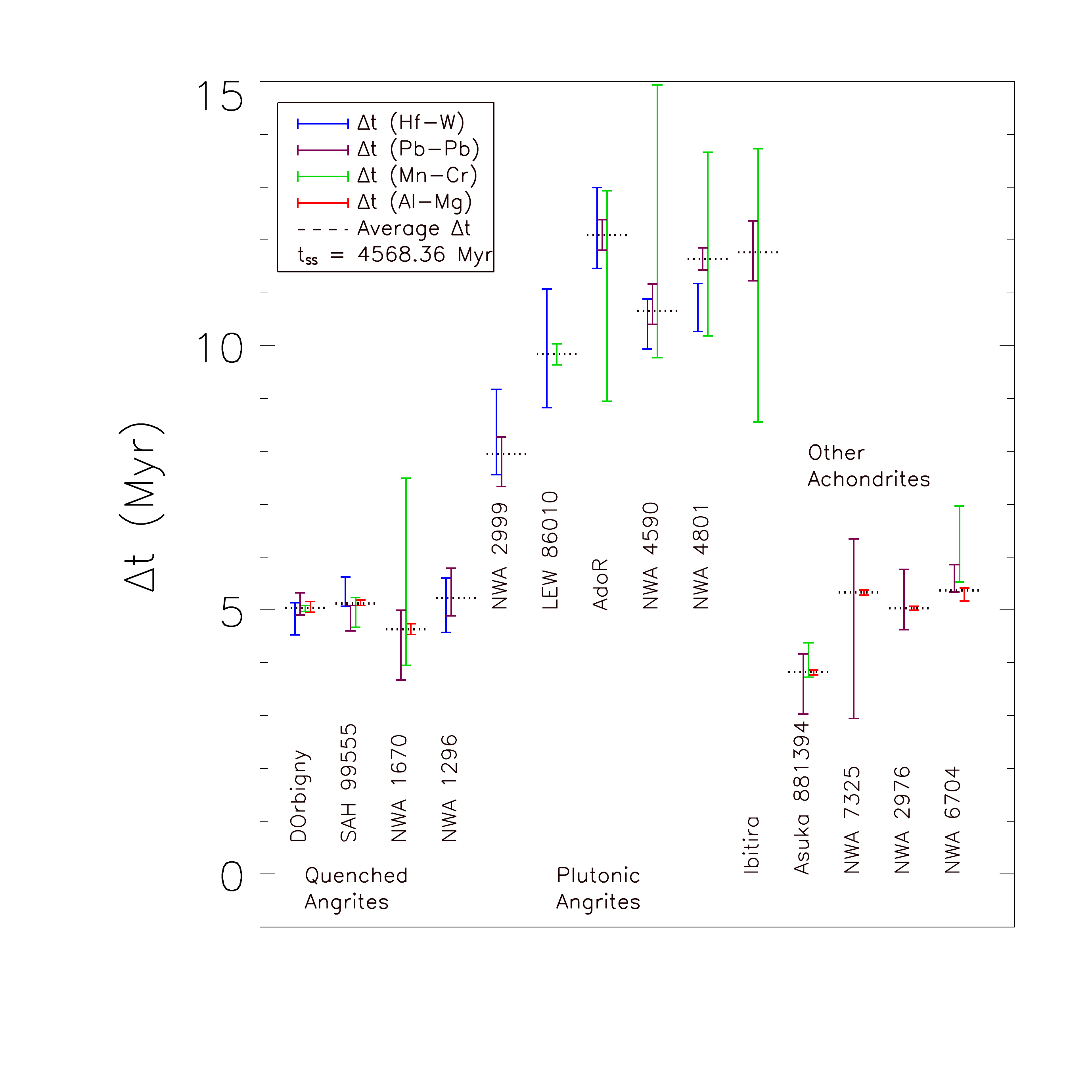}
\vspace{-0.2in}
\caption{
Times of formation of 14 achondrites, using the following parameters:
$t_{\rm SS} = 4568.36$ Myr, $(\mnratio)_{\rm SS} = 8.09 \times 10^{-6}$, 
$\mnfivethree$ half-life 3.80 Myr, and $(\hfratio)_{\rm SS} = 10.42 \times 10^{-5}$. 
Using the measurements reported in Table 1, we calculate for each achondrite the 
formation times after $t\!\!=\!\!0$: $\Delta t_{26}$ (red), $\Delta t_{53}$ 
(green), $\Delta t_{182}$ (blue), and $\Delta t_{\rm Pb}$ (violet), 
and their weighted mean $\Delta t$ (black dashed line), as listed in Table~\ref{table:two}.
If concordant, the formation times of 95\% (i.e., all but two) of the ages from all isotopic systems should match $\Delta t$ within $< 2\sigma$ uncertainty.
Only the Mn-Cr age of NWA 6704 ($2.4\sigma$), the Pb-Pb age of SAH 99555 ($2.9\sigma$) and the Hf-W age of NWA 4801 ($4.2\sigma$) are discordant. 
Including NWA 4801, $\chisq = 1.41$, which is still statistically significant (6\% probability), but we consider it disturbed and exclude it.
The 37 formation times across four isotopic systems, in 14 achondrites, are then made concordant in a statistical sense ($\chisq = 1.09$, 33\% probability; deviations distributed normally) using only 4 input parameters. 
This supports rather than falsifies the assumption of homogeneity of radionuclides.
\label{fig:achondrites}}
\end{figure}
\end{center}

\section{Discussion}

\subsection{Comparison to CAIs}

The parameters we advocate were not derived using any knowledge of CAIs whatsoever, except that {\Steve our decision to define} $(\alratio)_{\rm SS} = 5.23 \times 10^{-5}$ at $t\!\!=\!\!0$ {\Steve was informed by CAIs}.
Therefore, measurements of the $\mnfivethree$ half-life, or direct inferences of $(\mnratio)_{\rm SS}$, $(\hfratio)_{\rm SS}$ or $t_{\rm SS}$ from measurements of CAIs, provide a severe test of our model's predictions.

\subsubsection{$\mnfivethree$ half-life}

Our inferred half-life of $\mnfivethree$, {\Steve $3.80 \pm 0.23 (2\sigma)$ Myr, is much more precise} than the oft-quoted value $3.7 \pm 0.37 (1\sigma)$ Myr \citep{HondaImamura1971}.
[NB: It is common in the experimental physics literature to report uncertainties in measured values as the standard deviation in the data, i.e., as $1\sigma$ uncertainties. In contrast, it is common in the cosmochemistry literature to report uncertainties as 95\% confidence intervals, i.e., $2\sigma$ uncertainties. In citations below where the uncertainty was not defined, we add a question mark.]
In fact, most measurements of the $\mnfivethree$ half-life are quite uncertain. 
Besides the commonly cited value $3.7 \pm 0.37 (1\sigma)$ Myr, there are $2.9 \pm 1.2 (1\sigma?)$ Myr \citep{MatsudaEtal1971} and $3.9 \pm 0.6 (1\sigma?)$ Myr \citep{WoelfleEtal1973}. 
These three values are all within uncertainty of their weighted mean of  $3.70 \pm 0.61 (2\sigma)$ Myr.

There also is a long history of trying to reconcile Mn-Cr and other ages in meteorites,  leaving the $\mnfivethree$ half-life as a free parameter.
Using Apollo 14 samples to measure cosmogenic $\mnfivethree$, \citet{HerrEtal1972} derived 
$3.8 \pm 0.7 (1\sigma?)$ Myr. 
Reconciling Mn-Cr systematics with Al-Mg systematics in 18 chondrites, 
\citet{HeimannEtal1974} found 
$3.85 \pm 0.4 (1\sigma?)$ Myr.
Likewise, \citet{NyquistEtal2009} compared Al-Mg and Mn-Cr ages in achondrites and CAIs and inferred that the best fit to the $\mnfivethree$ half-life was that it was the $\altwosix$ half-life divided by $(0.23 \pm 0.04)$, implying a half-life $\approx 3.1$ Myr.
On the other hand, the regression they performed of Mn-Cr against Pb-Pb ages, using LEW 86010 and Asuka 881394, suggested a much longer half-life of 4.8 Myr.
\citet{SanbornEtal2019} 
regressed Mn-Cr ages against Pb-Pb relative ages for several achondrites, allowing the half-life of $\mnfivethree$ to be a free parameter, and found a decay constant $1.8 \pm 0.2 (2\sigma) \times 10^{-7} \, {\rm yr}^{-1}$, equivalent to a $3.85 \pm 0.43 (2\sigma)$ Myr half-life. 
{\Steve Our inferred value is exactly in line with these estimates, but more precise.}

\subsubsection{Initial ratio $(\mnratio)_{\rm SS}$}

Presumably, a measurement of $(\mnratio)_0$ in CAIs would directly record $(\mnratio)_{\rm SS}$, if the Mn-Cr system closed at the same time as the Al-Mg system.
However, the initial value $(\mnratio)_0$ in CAIs is very difficult to infer from measurements,
{\EDIT because Mn concentrations and Mn/Cr ratios are low in CAIs, among other reasons \citep{DavisMcKeegan2014}.}
For various CAIs, \citet{BirckAllegre1985} reported $(\mnratio)_0$ $= (44 \pm 10) \times 10^{-6}$, \citet{PapanastassiouEtal2005} reported $(\mnratio)_0 = (14.33 \pm 5.48) \times 10^{-6}$, \citet{NyquistEtal2009} reported $(\mnratio)_0 = (9.1 \pm 1.7) \times 10^{-6}$, and \citet{TrinquierEtal2008} reported $(\mnratio)_0 = (6.28 \pm 0.66) \times 10^{-6}$. 
The weighted average of the three more recent (and more precise) measurements is $(6.74 \pm 0.61) \times 10^{-6}$, and all three are marginally concordant with this value at the roughly $3\sigma$ level. 
Presumably this is the Solar System value that CAIs obtained when they formed, so $(\mnratio)_{\rm SS} \approx 6.74 \times 10^{-6}$, if the Mn-Cr system was not later reset.
Slightly higher values ($> 8 \times 10^{-6}$) might be inferred if isotopic closure of the Mn-Cr system in CAIs took place $\sim 1$ Myr after $t\!\!=\!\!0$. 
In analogy with the conclusions we reached in Paper I for the ability of the Pb-Pb system to be reset by transient heating of CAIs, such late resetting of the Mn-Cr system in CAIs may be likely.

In an approach similar to ours but more comprehensive, \citet{TissotEtal2017} recently reviewed various measurements to derive $(\mnratio)_0$ in bulk chondrites, plus anchoring to D’Orbigny to extrapolate backward in time to derive $(\mnratio)_{\rm SS}$.
Summarizing these, they recommended $(7 \pm 1) \times 10^{-6}$, with the Pb-Pb age of CAIs being a major uncertainty. 
However, if the Pb-Pb age of CAIs were fixed at $4567.94$ Myr
\citep{BouvierEtal2011a}, they state that they would then recommend a somewhat higher value, $(\mnratio)_{\rm SS} = (7.37 \pm 0.60) \times 10^{-6}$.
Our estimate, {\Steve $(8.09 \pm 0.65) \times 10^{-6}$} is in line with the high end of their estimate.

 \subsubsection{Initial ratio $(\hfratio)_{\rm SS}$}
 
The initial value $(\hfratio)_0$ in CAIs has been a challenge to measure, and only recently has it been very well constrained. 
\citet{BurkhardtEtal2012} reported $(\hfratio)_0 = (9.85 \pm 0.40) \times 10^{-5}$, based on a Hf-W isochron for mineral separates of a coarse-grained, type B Allende CAI. 
Because this CAI was melted after its minerals formed, the true initial value in the Solar System was likely higher; if melted $\sim 1$ Myr after $t\!\!=\!\!0$, $(\hfratio)_{\rm SS}$ could have been as high as $(10.65 \pm 0.43) \times 10^{-5}$.
Indeed, \citet{KruijerEtal2014} reported $(10.49 \pm 0.62) \times 10^{-5}$, based on their investigation of fine-grained (unmelted) CAIs. 
The value they recommend, $(10.18 \pm 0.43) \times 10^{-5}$, is based 
on a weighted average of fine-grained plus coarse-grained CAIs. 
Our value inferred from making ages concordant, {\Steve $(\hfratio)_{\rm SS} = (10.42 \pm 0.23) \times 10^{-5}$}, is in excellent agreement with the value of $(\hfratio)_0$ inferred from measurements of fine-grained CAIs \citep{KruijerEtal2014}.

\subsubsection{Pb-Pb age of the solar system, $t_{\rm SS}$}

The most surprising result from the above analysis 
{\Steve and of Paper I, in which $t_{\rm SS} = 4568.42 \pm 0.20$ Myr was estimated},
is that it predicts a Pb-Pb age of CAIs of {\Steve $4568.36 \pm 0.19 \, {\rm Myr}$} if the Pb-Pb system closed at the same time as the Al-Mg system in these CAIs.
As discussed already in Paper I, this is inconsistent with the measured Pb-Pb ages of CAIs, which are $4567.18 \pm 0.50$ for Allende CAI \textit{SJ101} \citep{AmelinEtal2010}, and $4567.35 \pm 0.28$ Myr, $4567.23 \pm 0.29$ Myr, and $4567.38 \pm 0.31$ Myr for CAIs \textit{22E}, \textit{31E}, and \textit{32E} from Efremovka \citep{ConnellyEtal2012}.
{\Steve Our inferred age of $t_{\rm SS}$ is over 1 Myr older than these CAI ages. However,}
some reports suggest older Pb-Pb ages of CAIs: $4567.94 \pm 0.31$ Myr for CAI \textit{B4} from NWA 6991 \citep{BouvierEtal2011a}, and $4568.22 \pm 0.18$ Myr for CAI \textit{B1} from NWA 2364 \citep{BouvierWadhwa2010}.
The former was not published in the refereed literature, the latter was not corrected using a direct measurement of the $\uratio$ ratio. 
{\Steve Interestingly, our inferred value of $t_{\rm SS}$ is consistent with these CAI ages.}

In Paper I we demonstrated that transient heating events with the peak temperatures and cooling rates characteristic of chondrule formation could have reset the Pb-Pb system in CAIs without resetting the Al-Mg system, allowing the Pb-Pb ages to look younger than the Al-Mg ages.
As CAIs resided in the protoplanetary disk until incorporated into chondrites, and heating of chondrules was ongoing for many Myr, CAIs could appear through their Pb-Pb ages to have formed up to several Myr after $t\!\!=\!\!0$.
This potentially explains the 1 Myr discrepancy between oft-cited Pb-Pb ages of CAIs and our inferred $t_{\rm SS}$.


\subsection{Homogeneity and concordancy in other samples}

Throughout this work, {\Steve based on the arguments presented in Paper I comparing the Al-Mg and Pb-Pb chronometers},
we have assumed homogeneity of the SLRs, especially $\altwosix$.
{\Steve This hypothesis would have been falsified if the formation times in rapidly cooled achondrites, especially quenched angrites, were discordant.
In Paper I we found that the Al-Mg and Pb-Pb formation times of volcanic achondrites were concordant.
Here we find that the Al-Mg and Pb-Pb formation times of all achondrites, and indeed the Hf-W and Mn-Cr formation times, are also concordant.
This finding further supports the assumption of homogeneity.}
Nevertheless, there are other samples for which simultaneous closure of multiple isotopic systems might be likely; these could potentially invalidate our model's assumption of SLR homogeneity. 
We examine some here, including FUN {\EDIT (Fractionations and Unknown Nuclear effects)} CAIs, chondrules and other components of CR chondrites, and CB/CH chondrites.

\subsubsection{Homogeneity of SLRs inferred from FUN CAI STP-1}

It has been suggested that the FUN CAI STP-1 provides evidence for decoupling of $\hfoneeighttwo$ and $\altwosix$ in the solar nebula \citep{HolstEtal2013,ParkEtal2017}. 
STP-1 exhibits $(\alratio)_0 = (2.94 \pm 0.21) \times 10^{-6}$ and $(\hfratio)_0 = (9.60 \pm 1.10) \times 10^{-5}$ if inferred using ${}^{186}{\rm W}/{}^{183}{\rm W}$ to correct for fractionation, or $(\hfratio)_0 = (9.22 \pm 1.10) \times 10^{-5}$ if using ${}^{186}{\rm W}/{}^{184}{\rm W}$ \citep{HolstEtal2013}.
The argument made by \citet{HolstEtal2013} is that the $(\alratio)_0$ ratio is far below the canonical value $(\alratio)_{\rm SS} = 5.23 \times 10^{-5}$, implying it formed a long time after CAIs (if $\altwosix$ was homogeneous), whereas its $(\hfratio)_0$ value is ``identical" within uncertainties to the initial Solar System $(\hfratio)_{\rm SS}$, which was taken to be $(9.85 \pm 0.40) \times 10^{-5}$ \citep{BurkhardtEtal2012}.
This would imply STP-1 formed at the same time as other CAIs but had only 6\% the canonical amount of $\altwosix$. 
Put another way, \citet{HolstEtal2013} inferred $\Delta t_{26} = 3.02 \pm 0.07$ Myr [based on a different $(\alratio)_{\rm SS}$], but $\Delta t_{182} = 0.33_{-1.47}^{+1.67}$ Myr, and considered this age gap of 2.69 Myr, discrepant at the $3.4\sigma$ level, to be irreconcilable.
In their interpretation, FUN CAIs formed in a region relatively devoid of $\altwosix$. 

This conclusion depends very strongly, however, on the assumed value of
$(\hfratio)_{\rm SS}$.
We and \citet{KruijerEtal2014} both infer much higher values than the previously accepted value of $(9.85 \pm 0.40) \times 10^{-5}$ of \citet{BurkhardtEtal2012}.
Using $(\hfratio)_{\rm SS} = 10.42 \times 10^{-5}$ and the 
${}^{186}{\rm W}/{}^{184}{\rm W}$-normalized value ($\hfratio)_{0} = (9.22 \pm 1.10) \times 10^{-5}$ yields $\Delta t_{182} = 1.57_{-1.45}^{+1.63}$ Myr.
{\Steve The discrepancy between this age and $\Delta t_{26} = 2.98 \pm 0.07$ Myr is only 1.41 Myr (1.7$\sigma$).
The $\Delta t_{26}$ and updated $\Delta t_{182}$ formation times do not differ by enough to exceed the usual threshold for discrepancy ($2\sigma$).}
We therefore conclude that the Hf-W and Al-Mg ages of STP-1 do not provide
strong evidence that $\altwosix$ and $\hfoneeighttwo$ were decoupled in 
the solar nebula.

\subsection{CR Chondrites and Chondrules} 

With our updates to the predicted Pb-Pb age of CAIs, $(\mnratio)_{\rm SS}$
and $(\hfratio)_{\rm SS}$, a valid concern is that other ``anchor" systems currently thought to be concordant no longer will be.
This may be particularly true for the ages of CR chondrites and their 
chondrules, because these have some of the largest times of formation after
$t\!\!=\!\!0$ of any chondrites (therefore small discrepancies have time to
accumulate). 
They also have been dated by Pb-Pb, Al-Mg and Hf-W systems, making them
vulnerable to changes in any one of these systems. 

Currently the ages of CR chondrite chondrules derived from different systems have appeared to be concordant with a formation time of about 3.7 Myr after CAIs. 
\citet{BuddeEtal2018} measured ${}^{182}{\rm W}$ excesses 
in CR chondrites. 
This dates the time of metal-silicate separation of the metal in CR chondrites which, they argue, was simultaneous with formation of the chondrules in CR chondrites. 
Combining several samples (bulk CR chondrites and bulk CR chondrules) into a single isochron, they derived 
$(\hfratio)_0 =  (7.68 \pm 0.17) \times  10^{-5}$. 
They combined this with an assumed 
$(\hfratio)_{\rm SS} = (10.18 \pm 0.43) \times 10^{-5}$ 
\citep{KruijerEtal2014} to derive a time of metal-silicate separation/chondrule formation in CR chondrites, $\Delta t_{182} = 3.63 \pm 0.62$ Myr. 
They also compared to Pb-Pb ages of six CR chondrite chondrules, analyzed by
\citet{AmelinEtal2002}, but corrected for their U isotopic compositions by
\citet{SchraderEtal2017}, who found 
$t_{\rm Pb} = 4563.6 \pm 0.6$ Myr. 
Using the Pb-Pb age of CAIs of $4567.30 \pm 0.16$ Myr of
\citet{ConnellyEtal2012},  they inferred 
$\Delta t_{\rm Pb} = 3.66 \pm 0.63$ Myr for these chondrules. 
\citet{BuddeEtal2018} also compared to a time 
$\Delta t_{26} = 3.75 \pm 0.24$ Myr of these chondrules’
formation they say is inferred by \citet{SchraderEtal2017} using Al-Mg
systematics. 
They noted these are concordant, implying that chondrules in CR chondrites
formed at about 3.7 Myr after CAIs. 
One could calculate a weighted mean $\Delta t = 3.73 \pm 0.21$ Myr after
CAIs, and all the systems would appear concordant with that time of
formation.

We interpret these results differently.
First, it is not clear that $\Delta t_{26} = 3.75 \pm 0.24$ Myr is appropriate. 
\citet{SchraderEtal2017} identified three groups of chondrules in CR chondrites: group 1 (n=1) had $(\alratio)_0 = (6.3 \pm 0.9) \times 10^{-6}$, group 2 (n=7) had $(\alratio)_0 = (3.2 \pm 0.6) \times 10^{-6}$, and group 3 (n=14) had $(\alratio)_0 = (6.7 \pm 0.3) \times 10^{-7}$.
\citet{SchraderEtal2017} took the weighted mean of the group 2 and group 3 chondrules’ $(\alratio)_0$ values, and then converted that average $(\alratio)_0$ to a time of formation assuming a $\altwosix$ half-life of 0.705 Myr, to report $\Delta t_{26} = 3.7_{-0.2}^{+0.3}$ Myr.  
From there, \citet{BuddeEtal2018} reported this as $\Delta t_{26} = 3.75 \pm 0.24$ Myr. 
However, when we take a weighted mean of the $(\alratio)_0$ values, we find it is $(6.71 \pm 0.08) \times 10^{-7}$, essentially the group 3 value, both because there are more group 3 chondrules, and because their $(\alratio)_0$ values are known to greater (absolute) precision.
Using this ratio, and a $\altwosix$ half-life 0.717 Myr, we the find times of formation of the CR chondrules to be {\Steve $\Delta t_{26} = 4.51 \pm 0.21$ Myr}. 
Next, using our determination of {\Steve $t_{\rm SS} = 4568.36 \pm 0.20$ Myr, we calculate $\Delta t_{\rm Pb} = 4.76 \pm 0.64$ Myr.}
These two times of formation, determined by internal isochrons, have weighted mean {\Steve $\Delta t = 4.53 \pm 0.20$ Myr}, with the Al-Mg and Pb-Pb ages concordant with each other. 
Finally, using ($\hfratio)_{\rm SS} = (10.42 \pm 0.20) \times 10^{-5}$, and a 
$\hfoneeighttwo$ half-life 8.90 Myr, we find {\Steve $\Delta t_{182} = 3.92 \pm 0.35$ Myr}. 
{\Steve
This latter age is slightly discordant ($2.9\sigma)$, but as it is based on an excess and not an internal isochron, it may reflect an early stages of metal-silicate separation while the chondrules were in the solar nebula, while the other two chronometers could record a slightly later heating event.}
We suggest that the CR chondrites may have assembled as late as {\Steve 4.5 Myr} after $t\!\!=\!\!0$. 

\subsection{Chondrules in CB/CH chondrites}

Another apparently concordant system involves the ages of chondrules in CB and CH chondrites.
Chondrules and CAIs in the CB and CH chondrites formed or were reset much later than similar inclusions in other chondrites, a fact attributed to their formation or resetting in an impact plume following a collision between two asteroids \citep{KrotEtal2005}.
Because of the suddenness of the event, it has been suggested \citep{BollardEtal2019} that this event could serve as a time anchor, possibly a better time anchor than the angrites.
Various chronometers have been applied to date the timing of the impact.

\subsubsection{Al-Mg ages of CB/CH chondrules}

So far, the precise Al-Mg chronometer has been used only to provide a lower limit to the time of formation. 
Very low $(\alratio)_0$ values $\sim 10^{-6}$ or lower in chondrules in CH or CB chondrites generally argue for a time of formation $\Delta t_{26} > 4$ Myr \citep{WeberEtal1995,KrotEtal2005}.
For example, \citet{OlsenEtal2013} found $(\alratio)_0 = (4.5 \pm 8.3) \times 10^{-7}$, implying $\Delta t_{26} > 3.8$ Myr, from bulk Hammada al Hamra 237 chondrules.

\subsubsection{Pb-Pb ages of CB/CH chondrules}

\citet{KrotEtal2005} derived Pb-Pb ages of chondrules from the CB chondrite Gujba, reporting an average age $4562.68 \pm 0.49$ Myr. 
\citet{BollardEtal2019} pointed out that the Pb-Pb ages reported by \citet{KrotEtal2005} were not U-corrected.
Gujba bulk chondrite has been measured at $\uratio = 137.794 \pm 0.014$ \citep{ConnellyEtal2012}, which matches the value advocated by \citet{GoldmannEtal2015} to be used for bulk Solar System, $137.794 \pm 0.027$. 
\citet{BollardEtal2015} instead applied a correction for $\uratio = 137.786 \pm 0.013$ that they asserted should be used for Solar System objects \citep{ConnellyEtal2012}, to derive an absolute age $4561.68 \pm 0.51$ Myr. 
If we apply the value measured for Gujba, $\uratio = 137.794 \pm 0.014$, we derive a U-corrected Pb-Pb age of $4561.77 \pm 0.54$ Myr. 

\citet{BollardEtal2015} went on to measure Pb-Pb ages of 4 chondrules from Gujba. 
Pooling these 4 chondrules together, assuming $\uratio = 137.786$ (not measured), 
they derived an age $4562.49 \pm 0.21$ Myr. 
[If we apply the value measured for Gujba, $\uratio = 137.794 \pm 0.014$, we derive a U-corrected Pb-Pb age of $4562.57 \pm 0.24$ Myr.] 
Based on a Pb-Pb age of CAIs of $4567.30 \pm 0.16$ Myr, \citet{BollardEtal2019} inferred a time of formation of $\Delta t_{\rm Pb} = 4.81 \pm 0.33$ Myr. 

The U-corrected Pb-Pb ages of chondrules from Gujba are $4561.77 \pm 0.54$ Myr (using data from \citet{KrotEtal2005}) and $4562.57 \pm 0.24$ Myr (using data from \citet{BollardEtal2019}). 
{\EDIT New data from} \citet{ConnellyEtal2021} {\EDIT suggest $4562.64 \pm 0.13$ Myr for a single Gujba chondrule.}
Assuming these all mark a simultaneous event, their weighted mean is 
{\EDIT $4562.60 \pm 0.11$ Myr, consistent with the 2005 data at the $3\sigma$ level, and the 2019 data at the $0.3\sigma$ level.}
Based on this age, and our estimate $t_{\rm SS} = 4568.36$ Myr, we 
derive a time of formation {\EDIT $\Delta t_{\rm Pb} = 5.76 \pm 0.11$ Myr}.

\subsubsection{Mn-Cr ages of CB/CH chondrules}

\citet{YamashitaEtal2010} reported an initial value $(\mnratio)_0 = (3.18 \pm 0.52) \times 10^{-6}$  in Gujba chondrules and a metal grain.  
They used a value $(\mnratio)_0 = (3.24 \pm 0.04) \times 10^{-6}$ \citep{GlavinEtal2004} and an absolute age (a Pb-Pb age not U-corrected) of $4564.42 \pm 0.12$ Myr \citep{Amelin2008a} for D’Orbigny, plus a half-life of $\mnfivethree$ of 3.7 Myr, to infer an absolute age $4564.3 \pm 0.9$ Myr. 
Alternatively, they used a value $(\mnratio)_0 = (1.25 \pm 0.07) \times 10^{-6}$ \citep{LugmairShukolyukov1998} and an absolute age (a Pb-Pb age not U-corrected) of $4558.55 \pm 0.15$ Myr \citep{Amelin2008a} for LEW 86010 to derive an absolute age $4563.5 \pm 1.1$ Myr. 
They note that the absolute age $4564.3 \pm 0.9$ Myr was not concordant with the Pb-Pb ages of Gujba chondrules reported by \citet{KrotEtal2005}, $4562.68 \pm 0.49$ Myr. 
It also would not be concordant with the updated age of \citet{BollardEtal2015}, $4561.68 \pm 0.51$ Myr. 
If these ages were anchored to a presumed Pb-Pb age of CAIs of $4567.30 \pm 0.16$ Myr \citep{ConnellyEtal2012}, it would imply a time of formation $\Delta t_{53} \approx 3.00 \pm 0.91$ Myr or $\Delta t_{53} \approx 3.80 \pm 1.11$ Myr, depending on which other anchor is used. 

Our approach is more straightforward. 
Assuming an initial Solar System ratio 
{\Steve $(\mnratio)_{\rm SS} = 8.09 \times 10^{-6}$
and a $\mnfivethree$ half-life 3.80 Myr}, we use the
measured value $(3.18 \pm 0.52) \times 10^{-6}$ \citep{YamashitaEtal2010} to derive {\Steve $\Delta t_{53} = 5.08 \pm 0.90$ Myr}.
{\Steve This uncertainty is driven by the uncertainty slope of the Mn-Cr isochron and would be the same whether anchored to D'Orbigny or referenced to the Solar System initial value.}

\subsubsection{Hf-W ages of CB/CH chondrules}

\citet{BollardEtal2015} reported an average value 
$\epsilon^{182}{\rm W} = -2.97 \pm 0.16$ for CB chondrites. 
Based on a comparison to the value $\epsilon^{182}{\rm W} = -3.49 \pm 0.07$
pertinent to average CAIs \citep{KruijerEtal2014},
they derived a time of formation $\Delta t_{182} = 5.09 \pm 0.59$ Myr. 
We argued above (\S 5.1.3) that it would more accurate to use the value for fine-grained CAIs, 
$\epsilon^{182}{\rm W} = -3.57 \pm 0.07$ \citep{KruijerEtal2014}, 
in which case we would derive a time of formation of CB chondrites 
$\Delta t_{182} = 5.87 \pm 0.68$ Myr.

\subsubsection{Summary}

Assuming the CB chondrites and chondrules formed or were reset together in
the impact plume, and achieved isotopic closure simultaneously, we take a
weighted mean of the ages above {\EDIT [ $\Delta t_{\rm Pb} = 5.76 \pm 0.11$ Myr}, $\Delta t_{53} = 5.08 \pm 0.90$ Myr, and $\Delta t_{182} = 5.87 \pm 0.68$ Myr to find a time of formation {\EDIT $\Delta t = 5.75 \pm 0.11$ Myr.} 
All three systems are concordant with an age of 5.75 Myr at the $0.2\sigma$ (Pb-Pb), $1.5\sigma$ (Mn-Cr), and $0.4\sigma$ (Hf-W) levels. 
Following the suggestion of \citet{BollardEtal2015}, the CB chondrites and chondrules do appear to be concordant and would serve as a good anchor, but with an absolute age {\EDIT $\approx 4562.60$ Myr and a much later time of formation of $\Delta t \approx 5.75$ Myr after CAIs}.


\subsection{Formation Times of ``Anchors"}

Our updated estimates of quantities like $t_{\rm SS}$ allow refinements in the calculation of when particular samples formed.
{\Steve
Useful among these are those objects for which three or more precisely measured isotopic systems appear to have reached isotopic closure simultaneously and are apparently most concordant. 
These include the volcanic angrites D'Orbigny, SAH 99555, and NWA 1670, 
plus the chondrules formed in the impact plume following the CB/CH impact.} For lack of a better word we call them anchors.
In {\bf Table~\ref{table:three}}, we list: the time of formation after $t\!\!=\!\!0$ of the object, as recorded by the different isotopic systems, assuming our canonical parameters derived from volcanic achondrites; the formation time as inferred from combining the systems; and the model age for the object.
We include the latter only to allow ease of comparison with other works, and caution that these should not be used to anchor to.
We emphasize again that the absolute age has little use in models of planet formation or protoplanetary disk evolution; the time of formation after $t\!\!=\!\!0$ is the much more relevant quantity.
These quantities should be determined by using $(\alratio)_{0}$, or $(\mnratio)_{0}$ in conjunction with the $(\mnratio)_{\rm SS}$ and $\tau_{53}$ derived here, or $(\hfratio)_{0}$ and the $(\hfratio)_{\rm SS}$ derived here, or a Pb-Pb age and the value of $t_{\rm SS}$ derived here. 
Where possible, multiple ages should be combined.

\begin{table}[h!]
\noindent
\begin{minipage}{6.0in}
\footnotesize
\caption{Our model predictions (in Myr, with uncertainties) for samples in which all isotopic systems closed simultaneously: 
 times of formation after $t\!\!=\!\!0$, according to the Al-Mg, Mn-Cr, Hf-W, and Pb-Pb systems; weighted mean times of formation; and absolute ages (using standard half-lives). {\EDIT The absolute ages are provide for the sake of comparison, but should not be used as anchors.} \label{table:three}}
\bigskip
\begin{tabular}{l|rc|rc|rc|rr|rc|rc}

\hline \hline \\
{\bf Sample} & $\Delta t_{26}$ & $2\sigma$ & 
$\Delta t_{53}$ & $2\sigma$ & 
$\Delta t_{182}$ & $2\sigma$ & 
$\Delta t_{\rm Pb}$ & $2\sigma$ &
$\Delta t$ & $2\sigma$ & 
$t \,\,\,\,\,\,\,\,$ & $2\sigma$ \\ \hline
{\bf D'Orbigny} & 5.06 & 0.10 & 5.03 & 0.06 & 4.83 & 0.31 & 5.11 & 0.21 & {\bf 5.03} & {\bf 0.05} & 4563.32 & 0.20 \\
{\bf SAH 99555} & 5.14 & 0.05 & 4.95 & 0.29 & 5.35 & 0.28 & 4.84 & 0.24 & {\bf 5.12} & {\bf 0.05} & 4563.23 & 0.20 \\
{\bf NWA 1670} & 4.64 & 0.10 & 5.72 & 1.78 & $\,$ & $\,$ & 4.33 & 0.66 & {\bf 4.63} & {\bf 0.10} & 4563.72 & 0.22 \\
{\bf Asuka 881394} & 3.82 & 0.04 & 4.04 & 0.32 & $\,$ & $\,$ & 3.59 & 0.53 & {\bf 3.82} & {\bf 0.04} & 4564.53 & 0.20 \\
{\bf CH/CB} & $> 3.8$ & $\,$ & 5.08 & 0.90 & {5.87} & {0.68} & {\EDIT 5.76} & {\EDIT 0.11} & {\EDIT {\bf 5.75}} & {\EDIT {\bf 0.11}} & {\EDIT 4562.10} & {\EDIT 0.23} \\
$\,\,$ {\bf Chondrules} & $\,$ & $\,$ & $\,$ & $\,$ & $\,$ & $\,$ & $\,$ & $\,$ & $\,$ & $\,$ & $\,$ & $\,$ \\ \hline
\end{tabular}
\end{minipage}
\end{table} 

Measurements to determine data missing from Table~\ref{table:three} would provide a severe test of our model. 
For NWA 1670, we predict $(\hfratio)_0 \approx 7.3 \times 10^{-5}$.
For the CH/CB chondrules, we predict an initial ratio
{\Steve $(\alratio)_0 \approx 2.0 \times 10^{-7}$.} 
For NWA 7325, we predict $(\mnratio)_0 \approx 3.1 \times 10^{-6}$ and 
$(\hfratio)_0 \approx 6.9 \times 10^{-5}$.
For LEW 86010, we predict a U isotopic ratio leading to a Pb-Pb age $\approx 4558.58$ Myr.

An interesting application of the improved chronometry is to test whether the similar volcanic angrites D'Orbigny and SAH 99555 formed at the same time.
According to the Al-Mg chronometer alone, SAH 99555 formed $0.076 \pm 0.115$ Myr after D'Orbigny.
After combining all the isotopic systems, we conclude that SAH 99555 formed $0.091 \pm 0.068$ Myr after D'Orbigny.
Our predicted formation times are closer together than previously inferred ($0.090$ vs. $0.076$ Myr), but also more likely to be distinct ($2.7\sigma$ vs. $1.3\sigma$).

\section{Other Isotopic Systems Used for Chronometry}
 
Our updates to parameters, especially $t_{\rm SS}$, allow refinements in the quantities required for other isotopic systems to be used for chronometry.
Based on their common or increasing use, we discuss the I-Xe, Fe-Ni, Pd-Ag, Nb-Zr, and Be-B systems.

\subsection{The I-Xe system}
 
The decay of ${}^{129}{\rm I}$ to ${}^{129}{\rm Xe}$ is one of the earliest used chronometers, but the technique to use this isotopic system differs from the others because of the way the measurements are made. 
An estimate of the initial ${}^{129}{\rm I}/{}^{127}{\rm I}$ ratio in a meteoritic sample can be obtained by irradiating the sample with neutrons, transmuting ${}^{127}{\rm I}$ to ${}^{128}{\rm I}$, which decays with a half-life of 25 minutes to ${}^{128}{\rm Xe}$.
Xe is then driven out of the sample and measured, and an isochron of ${}^{129}{\rm Xe}/{}^{132}{\rm Xe}$ vs.\ ${}^{128}{\rm Xe}/{}^{132}{\rm Xe}$ is constructed.
To the extent that the neutron absorption cross section of ${}^{127}{\rm I}$ is known, the correlation could provide the ${}^{129}{\rm I}/{}^{127}{\rm I}$ ratio. 
Because of uncertainty in the neutron cross section, it is difficult to measure the initial abundance of ${}^{129}{\rm I}/{}^{127}{\rm I}$ to an acceptable accuracy; however, it is possible to precisely measure the ratio of ${}^{129}{\rm I}/{}^{127}{\rm I}$ in a sample {\EDIT relative} to the initial ${}^{129}{\rm I}/{}^{127}{\rm I}$ in an anchor, usually taken to be the enstatite achondrite Shallowater. 
In this way, the relative age between the sample and Shallowater can be obtained.
To be useful for chronometry, we must determine the time after $t\!\!=\!\!0$ at which Shallowater formed, $\Delta t_{\rm SW}$.

Recently, \cite{PravdivtsevaEtal2017} and \citet{GilmourCrowther2017} summarized the concordancy between I-Xe and U-corrected Pb-Pb data of a variety of samples, and derived absolute ages of the Shallowater standard of $4562.4 \pm 0.2$ Myr and $4562.7 \pm 0.3$ Myr, respectively.
We favor the value of \citet{GilmourCrowther2017}, as it includes samples such as Ibitira and NWA 7325 not included by \citet{PravdivtsevaEtal2017}.
Using this age and {\Steve $t_{\rm SS} = 4568.36 \pm 0.10$ Myr}, we infer that Shallowater reached closure at {\Steve $\Delta t_{\rm SW} = 5.65 \pm 0.36$ Myr} after $t\!\!=\!\!0$.

It is also possible to accept chondrules from the CB/CH impact event as an anchor,
and take a time of formation {\EDIT $\Delta t = 5.75 \pm 0.11$ Myr}, then apply the correction by \citet{PravdivtsevaEtal2017}, that Shallowater formed $0.29 \pm 0.16$ Myr before that, to infer {\Steve $\Delta t_{\rm SW} = 5.45 \pm 0.19 \, {\rm Myr}$}.
{\EDIT It is encouraging that these two approaches---one based on I-Xe age difference anchored to the CB/CH impact event constrained by Mn-Cr, Hf-W and Pb-Pb ages; the other based on comparing I-Xe formation times with Pb-Pb ages---yield identical formation times within the certainties (at the 0.9$\sigma$ level).}
Taking the weighted mean of these then yields our best estimate {\EDIT for when Shallowater formed}: 
{\EDIT
\begin{center}
\mbox{\boldmath$\Delta t_{\rm SW} = 5.50 \pm 0.17 \, {\rm Myr}$}
\end{center}
}

For completeness, we note that the half-life of ${}^{129}{\rm I}$, which has long been cited as $15.7 \pm 0.6(1\sigma)$ Myr \citep{Emery1972}, has been updated.
\citet{PravdivtsevaEtal2017} reviewed different values in the literature and compared them to a value derived from regressing I-Xe data, concluding that a value near 16 Myr appeared correct. 
More recently, in a concerted effort using multiple measurement techniques, \citet{GarciaToranoEtal2018} determined the half-life to be \mbox{\boldmath$16.14 \pm 0.12(1\sigma)$}, the value adopted here.

Using a formation time {\EDIT $\Delta t_{\rm SW} = 5.50 \pm 0.17$ Myr} and the half-life of ${}^{129}{\rm I}$, the initial ${}^{129}{\rm I}/{}^{127}{\rm I}$ ratio can be estimated. 
In recent experimental work to effectively derive the neutron cross sections, \citep{PravdivtsevaEtal2021} estimated $({}^{129}{\rm I}/{}^{127}{\rm I})_0 \approx 1.35 \times 10^{-4}$ in Shallowater at the time of its formation, with an apparent uncertainty $< 2\%$. 
Extrapolating backward in time yields
\begin{center}
\mbox{\boldmath$({}^{129}{\rm I}/{}^{127}{\rm I})_{\rm SS} \approx (1.71 \pm 0.02) \times 10^{-4}$}.
\end{center}
{\EDIT This value is somewhat higher than estimates of $1.4 \times 10^{-4}$ \citep[e.g.,][]{Davis2022} because those are based on a value $1.07 \times 10^{-4}$ in Bjurb\"{o}le whole rock \citep{HohenbergKennedy1981}, which \citet{PravdivtsevaEtal2021} argue is too low because of self-shielding effects in the KI salts used in those experiments.}

\subsection{The Fe-Ni system}

The short-lived radionuclide ${}^{60}{\rm Fe}$ decays to ${}^{60}{\rm Ni}$ (via ${}^{60}{\rm Co}$), with a half-life of \mbox{\boldmath$2.62 \pm 0.04(1\sigma) \, {\rm Myr}$} \citep{RugelEtal2009}.
Before 2009, the half-life $1.49 \pm 0.27(1\sigma)$ Myr \citep{KutscheraEtal1984} was commonly cited, but the measurement by \citet{WallnerEtal2015}, $2.50 \pm 0.12(1\sigma)$ Myr supports the newer, more precise value.

The initial $({}^{60}{\rm Fe}/{}^{56}{\rm Fe})_{\rm SS}$ ratio was at first determined to be $\sim 10^{-6}$ \citep{TachibanaHuss2003}, but subsequent work has shown that this and similar analyses were overestimates, due to use of a too-short half-life, errors in how low ion counts obtained by Secondary Ion Mass Spectrometry (SIMS) analyses were averaged \citep{OglioreEtal2011,TelusEtal2013}, and the possibility that samples have suffered from redistribution of Fe and/or Ni isotopes after crystallization of samples \citep{QuitteEtal2011,TelusEtal2018}.
An analysis by sensitive Resonance Ionization Mass Spectrometery (RIMS) also has shown that SIMS analyses can introduce isotopic fractionation, leading to overestimates in $(\feratio)_0$ \citep{TrappitschEtal2018}.
A higher value $(\feratio)_{\rm SS} \approx 6 \times 10^{-7}$ inferred by \citet{CookEtal2021}, based on measured $\epsilon^{60}{\rm Ni}$ excesses in iron meteorites, is based on the assumption that the iron meteorite parent body had $\epsilon^{60}{\rm Ni} \approx 0.0$ like CI carbonaceous chondrites; assuming it was more CV chondrite-like (as implied by the $\epsilon^{62}{\rm Ni}$ excesses) yields $\epsilon^{60}{\rm Ni} \approx -0.13$ and admits $(\fesixty)_0 < 10^{-8}$.
Based on a variety of analyses yielding internal isochrons for CAIs, chondrules, achondrites and iron meteorites, especially by inductively coupled plasma mass spectrometry (ICP-MS), a value $({}^{60}{\rm Fe}/{}^{56}{\rm Fe})_{\rm SS} \sim 10^{-8}$ is now widely accepted \citep{QuitteEtal2010,SpivakBirndorfEtal2011,TangDauphas2012,TangDauphas2015}.

The initial $(\feratio)_{\rm SS}$ of the solar system can be found by extrapolating backwards from samples in which the Fe-Ni system closed at the same time as other systems. 
Using $\epsilon^{53}{\rm Cr}$ excesses to create a bulk rock isochron for the eucrite parent body (EPB), \citet{TrinquierEtal2008} determined $(\mnratio)_{0} = (4.21 \pm 0.42) \times 10^{-6}$, which implies {\Steve $\Delta t = 3.58 \pm 0.55$ Myr} for our favored parameters.
\citet{TangDauphas2012} inferred from bulk rock isochrons that $(\feratio)_0 = (3.45 \pm 0.32) \times 10^{-9}$ in the EPB, so extrapolating backwards we infer {\Steve $(\feratio)_{\rm SS} = (8.90 \pm 1.54) \times 10^{-9}$}.

{\Steve
Likewise, from whole rock isochrons of angrites, \citet{ShukolyukovLugmair2007} inferred $(\mnratio)_{0} = (3.40 \pm 0.14) \times 10^{-6}$ in the angrite parent body (APB), and \citet{ZhuEtal2019} $(3.16 \pm 0.11) \times 10^{-6}$, which together imply $\Delta t = 5.00 \pm 0.15$ Myr} for our favored parameters.
Combining several bulk-rock angrites, \citet{QuitteEtal2010} determined $(\feratio)_0 = (3.12 \pm 0.78) \times 10^{-9}$, and \citet{TangDauphas2012} inferred $(2.20 \pm 1.16) \times 10^{-9}$, with weighted mean $(\feratio)_0 = (2.83 \pm 0.65) \times 10^{-9}$.
Extrapolating backward, we infer {\Steve $(\feratio)_{\rm SS} = (10.61 \pm 2.47) \times 10^{-9}$.}
The weighted mean of these two analyses yields {\Steve $(\feratio)_{\rm SS} = (9.38 \pm 1.31) \times 10^{-9}$.}

It is also possible to extrapolate backward from the individual internal isochrons from D'Orbigny: $(\feratio)_0 = (4.1 \pm 2.6) \times 10^{-9}$ \citep{QuitteEtal2010}; $(2.81 \pm 0.86) \times 10^{-9}$ \citep{SpivakBirndorfEtal2011}; and $(3.42 \pm 0.58) \times 10^{-9}$ \citep{TangDauphas2012}; weighted average $(3.26 \pm 0.47) \times 10^{-9}$.
Using our inferred $\Delta t = 5.06 \pm 0.04$ Myr, we would infer $(\feratio)_{\rm SS} = (12.43 \pm 1.81) \times 10^{-9}$.
Likewise, we could use data for SAH 99555: $(\feratio)_0 = (1.8 \pm 0.5) \times 10^{-9}$ \citep{QuitteEtal2010}; $(1.97 \pm 0.77) \times 10^{-9}$ \citep{TangDauphas2012}.
With our inferred $\Delta t = 5.12 \pm 0.05$ Myr, we derive $(\feratio)_{\rm SS} = (7.17 \pm 1.64) \times 10^{-7}$. 
The weighted mean of these is $(9.5 \pm 1.2) \times 10^{-9}$, very close to the value inferred from bulk rock isochrons; but neither the D'Orbigny nor SAH 99555 {\EDIT internal isochrons are} concordant with that value (nor with each other). 

We consider the bulk rock isochrons more reliable, as they are less susceptible to thermal disturbance \citep{TangDauphas2012}.
The weighted mean of the values inferred from the EPB and APB is
\begin{center}
\boldmath$({}^{60}{\rm Fe} / {}^{56}{\rm Fe})_{\rm SS} = (9.4 \pm 1.3) \times 10^{-9}$. 
\end{center}
This is to be compared to the value $(11.5 \pm 2.6) \times 10^{-9}$ derived by \citet{TangDauphas2012}.
These are consistent with each other, and the difference is attributable mostly to our refinements in the half-life and abundance of $\mnfivethree$.

\subsection{The Pd-Ag system}

The SLR ${}^{107}{\rm Pd}$ decays to ${}^{107}{\rm Ag}$ with a half-life of \mbox{\boldmath$6.50 \pm 0.3(1\sigma) \, {\rm Myr}$} \citep{FlynnGlendenin1969}.
The Pd-Ag system has been used as a chronometer in the metallic phases of iron meteorites and mesosiderites, as well as in carbonaceous chondrites, and it would be useful to better constrain the initial ratio $({}^{107}{\rm Pd}/{}^{108}{\rm Pd})_{\rm SS}$ in the Solar System.
Isochrons in a variety of iron and stony-iron meteorites yield a range of initial $({}^{107}{\rm Pd}/{}^{108}{\rm Pd})_0 \approx (1.5 - 2.5) \times 10^{-5}$ \citep{ChenWasserburg1990, ChenEtal2002}.
Later studies found, after underestimating the ages of some iron meteorites, that the I{\sc ab} iron meteorite parent body likely suffered partial melting of metal and sulfides some 15 Myr after Solar System formation, resetting the Pd-Ag system \citep{TheisEtal2013}. 
Carbonaceous chondrites are therefore a better sample for constraining $({}^{107}{\rm Pd}/{}^{108}{\rm Pd})_{\rm SS}$.
\citet{SchonbachlerEtal2008} found $(\pdratio)_{\rm SS} = (5.9 \pm 2.2) \times 10^{-5}$ from whole-rock isochrons of carbonaceous chondrites. This was refined to a value $(\pdratio)_{\rm SS} = (6.6 \pm 0.4) \times 10^{-5}$ 
by \citet{MatthesEtal2018} and \citet{BrenneckaEtal2018} using measurements of
the type IV{\sc a} iron meteorite Muonionalusta, and assuming a Pb-Pb age of the Solar System 4567.3 Myr.

We refine this by taking the reported age for Muonionalusta, $4558.4 \pm 0.5$ Myr \citep{BrenneckaEtal2018}, to estimate a time of formation {\Steve $\Delta t = 9.96 \pm 0.61$ Myr.}
We then extrapolate backwards from the measured value $({}^{107}{\rm Pd}/{}^{108}{\rm Pd})_0 = (2.57 \pm 0.07) \times 10^{-5}$ \citep{MatthesEtal2018}, to derive 
\begin{center} 
\mbox{\boldmath$({}^{107}{\rm Pd}/{}^{108}{\rm Pd})_{\rm SS} = (7.43 \pm 0.52) \times 10^{-5}$}.
\end{center}
Including the $2\sigma$ uncertainty in the half-life would increase the uncertainty to $\pm (0.90) \times 10^{-5}$.

\subsection{The Nb-Zr system}

The short-lived radionuclide ${}^{92}{\rm Nb}$ decays to ${}^{92}{\rm Zr}$ with a half-life that is \mbox{\boldmath$34.7 \pm 2.4 \, {\rm Myr}$} \citep{AudiEtal2003}, although \citet{IizukaEtal2016} recommended using the older value $37 \pm 5$ Myr \citep{Holden1990}.

An initial ratio $({}^{92}{\rm Nb}/{}^{93}{\rm Nb})_0 = (1.4 \pm 0.5) \times 10^{-5}$ was recently determined for NWA 4590 \citep{IizukaEtal2016}.
Using a Pb-Pb age $4557.93 \pm 0.36$ Myr and $t_{\rm SS} = 4567.3$ Myr (i.e., $\Delta t = 9.37$ Myr), they extrapolated backward in time to infer
\begin{center}
\mbox{\boldmath$({}^{92}{\rm Nb}/{}^{93}{\rm Nb})_{\rm SS} = (1.7 \pm 0.6) \times 10^{-5}$}.
\end{center}
Although we infer $\Delta t = 10.64 \pm 0.30$ Myr for NWA 4590, the difference is negligible because of the long mean-life of ${}^{92}{\rm Nb}$.
The initial $({}^{92}{\rm Nb}/{}^{93}{\rm Nb})_{\rm SS}$ ratio in the Solar System is insensitive to these differences, but conversely the ages derived from the Nb-Zr are very sensitive to the $({}^{92}{\rm Nb}/{}^{93}{\rm Nb})_{\rm SS}$ ratio, so we recommend further determinations of this value to develop the Nb-Zr system as a chronometer. 

\subsection{The Be-B System}

The short-lived radionuclide ${}^{10}{\rm Be}$ decays to ${}^{10}{\rm B}$ with a half-life of 1.39 Myr.
\citet{ChmeleffEtal2010} reported an experimentally derived half-life 
$1.386 \pm 0.016 (1\sigma)$ Myr. 
Using a different experimental technique, 
\citet{KorschinekEtal2010} found $1.388 \pm 0.018 (1\sigma)$ Myr.
Both papers’ authors recommended combining their results and 
using the half-life \mbox{\boldmath$1.387 \pm 0.012 (1\sigma)$} {\bf Myr}.
The Be-B system has almost exclusively been studied in CAIs containing the minerals melilite, hibonite, and grossite, because these minerals are among the few to exhibit the required variable-to-high Be/B ratios (up to a few hundred) to build an isochron \citep{DunhamEtal2020}, and these minerals are unique to CAIs.
Identification of phases with variable Be/B in other meteoritic components may allow dating of other samples, and the Be-B system may serve as a good chronometer of events affecting CAIs, so it is worthwhile to establish $(\beratio)_{\rm SS}$.

Recently, \citet{DunhamEtal2022} determined that the $({}^{10}{\rm Be}/{}^{9}{\rm Be})_0$ ratios recorded by CAIs overwhelmingly cluster, with few exceptions, around a single value, 
\begin{center}
\mbox{\boldmath$({}^{10}{\rm Be}/{}^{9}{\rm Be})_{\rm SS} = 
(7.1 \pm 0.2) \times 10^{-4}$}.
\end{center}

Of the 54 $\beratio$ robust CAI regressions, 10 (19\%) have $({}^{10}{\rm Be}/{}^{9}{\rm Be})_0$ above (n=3) and below (n=7) the $({}^{10}{\rm Be}/{}^{9}{\rm Be})_{\rm SS}$. 
Overall, though, the homogeneity of ${}^{10}{\rm Be}$ in CAI regressions (81\%) suggests that ${}^{10}{\rm Be}$ was distributed uniformly in the solar nebula \citep{DunhamEtal2022}.

Most $({}^{10}{\rm Be}/{}^{9}{\rm Be})_0$ measurements were conducted on normal CAIs that have nearly canonical $(\alratio)$ ratio and $({}^{10}{\rm Be}/{}^{9}{\rm Be})_{\rm SS}$.
Of the seven CAIs with low $({}^{10}{\rm Be}/{}^{9}{\rm Be})_0 \approx (3-5)\times 10^{-4}$, all have isotopic anomalies (i.e., are FUN {\EDIT or PLAty Crystals of hibonite [PLAC]} type CAIs) and five are dominated by hibonite.
We consider it likely that hibonite-dominated particles did not sample the overall solar nebula reservoir, instead retaining memory of a presolar origin, a possibility previously suggested by \citet{Ireland1990}, \citet{KoopEtal2016}, and \citet{LarsenEtal2020}.
The only non-hibonite-dominated objects with measured and non-canonical $(\alratio)_0$ and $(\beratio)_0$ ratios are the FUN CAIs \textit{CMS-1} \citep{WilliamsEtal2017,DunhamEtal2022} and \textit{KT1} \citep{LarsenEtal2011,WielandtEtal2012}.
\textit{CMS-1} was a forsterite-bearing inclusion before thermal processing \citep{MendybaevEtal2017}, and records $({}^{10}{\rm Be}/{}^{9}{\rm Be})_0 = (1.8 \pm 3.2) \times 10^{-4}$ \citep{DunhamEtal2022} and $(\alratio)_{0} = (2.85 \pm 0.57) \times 10^{-5}$ \citep{WilliamsEtal2017}.
These imply $\Delta t_{10} = (2.75_{-2.05}^{+\infty}$ Myr (i.e., $\Delta t_{10} > 0.70$ Myr) and $\Delta t_{26} = (0.63_{-0.19}^{+0.23})$ Myr.
If the Al-Mg and Be-Be systems were simultaneously reset at $\Delta t = 0.8$ Myr, the predicted values for this inclusion would be $(\alratio)_0 = 2.4 \times 10^{-5}$ and $(\beratio)_0 = 4.8 \times 10^{-4}$, both within $< 2\sigma$ of the measured values.
CAI \textit{KT1} 
records $({}^{10}{\rm Be}/{}^{9}{\rm Be})_0 = (5.0 \pm 0.4) \times 10^{-4}$ \citep{DunhamEtal2022} and $(\alratio)_{0} = (-2.2 \pm 4.7) \times 10^{-5}$ \citep{LarsenEtal2011}.
These imply $\Delta t_{10} = (0.75 \pm 0.15)$ Myr and $\Delta t_{26} > 0.76$ Myr.
Interestingly, if the thermal processing events experienced by both CAIs took place at around $\Delta t = 0.8$ Myr, 
the Al-Mg and Be-B systems might be reconciled. 
More precise data for \textit{CMS-1}, \textit{KT1} and more non-hibonite-dominated FUN CAIs is needed to further test whether the Be-B system can be used as a chronometer of CAI melting events. 

\section{Conclusions}


{\EDIT
In Paper I we tested the hypothesis that $\altwosix$ was distributed homogeneously in the solar nebula, first finding the value of the Pb-Pb age of $t\!\!=\!\!0$ that minimized the discrepancies between the formation times $\Delta t_{26}$ found using Al-Mg measurements, and formation times $\Delta t_{\rm Pb}$ found using Pb-Pb dating.
We then tested whether this optimal fit made the ages concordant in a statistically sense. }
{\Steve 
For seven rapidly cooled achondrites, this was the case.
They could have falsified the homogeneity hypothesis, but did not.
This fact, plus the astrophysical theories for the origins of the short-lived radionuclides, strongly suggest all radionuclides were homogeneously distributed.}

{\Steve
In this paper we built on that model, creating further tests of homogeneity by comparing ages derived using the Hf-W and Mn-Cr systems as well.
For 11 achondrites (excluding only NWA 4801) with 26 formation times across the Al-Mg, Hf-W and Pb-Pb systems, we found statistical concordance using only two free parameters: $t_{\rm SS} = 4568.36$ Myr and $(\hfratio)_{\rm SS} = 10.43 \times 10^{-5}$. 
The goodness-of-fit parameter was 0.88, and the deviations in ages were normally distributed.
These findings {\bf strongly} support the assumption of homogeneity of $\altwosix$ and $\hfoneeighttwo$. 
They further suggest that the Hf-W system in NWA 4801 was lightly disturbed, which is consistent with the late thermal annealing inferred for this achondrite \citep{IrvingKuehner2007,McKibbinEtal2015} and the} {\EDIT high abundance of W in its matrix \citep{KleineEtal2012}.}

Extending the results to plutonic angrites and other achondrites, and including the Mn-Cr system, we found further concordancy. 
For preferred values of   
{\EDIT $\mnfivethree$ half-life 3.80 Myr, $(\mnratio)_{\rm SS}$ $=  8.09 \times 10^{-6}$, $(\hfratio)_{\rm SS} $ $= 10.42 \times 10^{-5}$, and $t_{\rm SS} = 4568.35$ Myr, we find 37 formation times across 14 achondrites are concordant in a statistical sense, with normally distributed $z$ scores and a goodness-of-fit parameter $\chisq \approx 1.1$.}
{\Steve This is our preferred solution.}

{\Steve
Our parameters were not based on any measurements of CAIs, other than to choose to reference $t\!\!=\!\!0$ to the time when $(\alratio) = 5.23 \times 10^{-5}$, because many CAIs have initial $(\alratio)_0$ close to this value.
Despite this, our values are in excellent accord with the few and imprecise measurements {\EDIT of $(\mnratio)_0$ and $(\hfratio)_0$} in CAIs. 
Although our value of $t_{\rm SS}$ is about 1 Myr older than the measured value of $t_{\rm Pb}$ in four CAIs \citep{ConnellyEtal2012}, it is in good accord with the two values inferred by \citet{BouvierWadhwa2010} and \citet{BouvierEtal2011a}.
We infer that late transient heating events like those undergone by chondrules can reset the Pb-Pb chronometer in CAIs without resetting Al-Mg, as discussed in Paper I.
}

{\Steve
Our results allow us to test the concordance of other isotopic systems such as chondrules forming after the impact event that created the CB/CH chondrites. 
We concur with \citet{BollardEtal2019} that the isotopic systems were likely to have closed simultaneously in these objects, making them a good test of our chronometry.
We find that if the impact took place at $5.75 \pm 0.11$ Myr, the Mn-Cr, Hf-W and Pb-Pb systems are indeed concordant, {\it using the same parameters derived from achondrites}.
}

{\Steve 
Our results allow other isotopic systems to be examined.
In particular, we conclude that Shallowater closed at $\Delta t_{\rm SW} = 5.50 \pm 0.23$ Myr after $t\!\!=\!\!0$.
This date should make it easier to put I-Xe ages into a solar nebula chronology.
{\EDIT I-Xe ages appear concordant.}
We do not see an obvious conflict between the Be-B and Al-Mg ages of the FUN CAIs \textit{CMS-1} and \textit{KT1}; both systems appear consistent with having formed at 0.8 Myr after $t\!\!=\!\!0$.
Our results provide context for studying these other isotopic systems.
}

{\Steve 
To assist with developing a sequence of events in the solar nebula, we strongly advocate reporting formation or closure times relative to $t\!\!=\!\!0$. 
There are no astrophysical models of planet or formation that would be affected if the solar nebula formed at 4500 or 4700 Myr ago instead of 4568 Myr.
Absolute ages are anyway uncertain to within $\pm 9$ Myr because of the uncertainties in the uranium half-lives; Pb-Pb dating is only precise when taking the difference between two ages, so that these uncertainties cancel.
In other words, Pb-Pb dating is only really useable to models and practically precise when it is employed as a relative chronometer. 
The use of anchors to report Al-Mg or Mn-Cr ages as modeled absolute ages is therefore not necessary, and indeed introduces considerable confusion and additional imprecision.
Our hope is that by constraining the initial values of $(\mnratio)_{\rm SS}$ and $(\hfratio)_{\rm SS}$ more precisely, this will enable reporting of relative ages.
}

{\Steve
The framework we have presented here makes clear the need for further measurements, and points to how to employ them.
From objects like D'Orbigny it is clear that pooling results from different laboratories has allowed dating that is more accurate and more precise; there is a need to measure all samples multiple times in different labs.
Pooling together data from multiple isotopic systems also leads to greater precision. 
For example, our inferred Al-Mg formation time of D'Orbigny is $\Delta t_{26} = 5.059 \pm 0.103$ Myr, but after combining all the data, our inferred time for formation is $\Delta t = 5.034 \pm 0.048$ Myr. 
This is twice as precise than the typical $> 0.1$ Myr uncertainty in Al-Mg formation times.
As more data are acquired for individual samples, more precise formation times can be inferred; and as more samples and data are acquired, our framework will allow more precise estimates of key quantities like $(\hfratio)_{\rm SS}$ and $(\mnratio)_{\rm SS}$. 
Already our uncertainties in these quantities are far less than those of single direct measurements of CAIs (for which disturbance cannot be ruled out anyway).
Our approach allows determination of ages and formation times of meteorites and inclusions, without the use of} {\EDIT individual samples as anchors}.
}

\bigskip
{\bf Acknowledgments}: The authors would like to acknowledge the efforts of cosmochemists from multiple laboratories around the world whose work makes possible the data cited in Table 1 and throughout this paper. Statistical chronometry necessarily distills very difficult and painstaking analytical work into mere numbers to be crunched, but the efforts to obtain those numbers are appreciated. 
We thank Zack Torrano for useful discussions. {\EDIT We thank Francois Tissot and two anonymous reviewers whose} suggestions greatly improved the quality of our work.
The work herein benefitted from collaborations and/or information exchange within NASA's Nexus for Exoplanetary System Science research coordination network sponsored by NASA's Space Mission Directorate (grant NNX15AD53G, PI Steve Desch).
Emilie Dunham gratefully acknolwedges support from a 51 Pegasi b Fellowship, grant \#2020-1829.

\bigskip

The data and scripts used to create Table 1 andFigure~\ref{fig:achondrites} are included as Research Data. 

\bigskip

\appendix

\section{Derivation of $t^{*}_{\rm SS}$}

\noindent
Here we derive Equation~\ref{eq:tss} for $t^{*}_{\rm SS}$.
The global goodness-of-fit parameter is 
\[
\!\!\!\!\!\!\!\!\!\!\!\!\!\!\!\!\!\!\!\!\!\!\!
\!\!\!\!\!\!\!\!\!\!\!\!\!\!\!\!\!\!\!\!\!\!\!
\!\!\!\!\!\!\!\!\!\!\!\!\!\!\!\!\!\!\!\!\!\!\!
\!\!\!\!\!\!\!\!\!\!\!\!\!\!\!\!\!\!\!\!\!\!\!
\chi_{\nu}^{2} = \frac{1}{N-M} \, \sum_{i=1}^{A} \left[ \frac{ \left( \Delta t_{26,i} - \Delta t_{i} \right)^2 }{ \sigma_{\Delta t26,i}^2 } \right.
\]
\begin{equation}
 \left. +\frac{ \left( \Delta t_{53,i} - \Delta t_{i} \right)^2 }{ \sigma_{\Delta t53,i}^2 } +\frac{ \left( \Delta t_{182,i} - \Delta t_{i} \right)^2 }{ \sigma_{\Delta t182,i}^2 } +\frac{ \left( \Delta t_{{\rm Pb},i} - \Delta t_{i} \right)^2 }{ \sigma_{\Delta t{\rm Pb},i}^2 } \right],
\end{equation}
and the optimal value of $t_{\rm SS}$ is that value which makes $\partial \chisq / \partial t_{\rm SS} = 0$.
It is recognized that $\partial (\Delta t_{{\rm Pb},i}) / \partial t_{\rm SS} = 1$ and, from the definition of $\Delta t_{i}$ (Equation~\ref{eq:deltat}), $\partial \Delta t_{i} / \partial t_{\rm SS} = \alpha_{i}$, where 
\begin{equation}
\alpha_i = \frac{
1 / \sigma_{\Delta t {\rm Pb},i}^2
}
{
1 / \sigma_{\Delta t {\rm Pb},i}^2
+ 1 / \sigma_{\Delta t26, i}^2
+ 1 / \sigma_{\Delta t53, i}^2
+ 1 / \sigma_{\Delta t182, i}^2
}
\end{equation}
Therefore 
\[
2 \sum_{i=1}^{A} \frac{ \Delta t_{i} - \Delta t_{26,i} }{ \sigma_{\Delta t 26,i}^2} \, \alpha_{i}
+2 \sum_{i=1}^{A} \frac{ \Delta t_{i} - \Delta t_{53,i} }{ \sigma_{\Delta t 53,i}^2} \, \alpha_{i}
+2 \sum_{i=1}^{A} \frac{ \Delta t_{i} - \Delta t_{182,i} }{ \sigma_{\Delta t 182,i}^2} \, \alpha_{i}
\]
\begin{equation}
+2 \sum_{i=1}^{A} \frac{ \Delta t_{i} - \Delta t_{{\rm Pb},i} }{ \sigma_{\Delta t{\rm Pb},i}^2 } \, \left( \alpha_{i} - 1 \right) = 0.
\end{equation}
All the terms involving $\alpha_{i}$ cancel (by construction, since $\Delta t_i$ itself was chosen to minimize $\chisq$), leaving only 
\begin{equation}
\sum_{i=1}^{A} \frac{ \Delta t_{{\rm Pb},i} }{ \sigma_{\Delta t{\rm Pb},i}^2 } = 
\sum_{i=1}^{A} \frac{ \Delta t_{i} }{ \sigma_{\Delta t{\rm Pb},i}^2 } .
\end{equation}
Substituting $\Delta t_{{\rm Pb},i} = t_{\rm SS} - t_{{\rm Pb},i}$, we can write this as 
\begin{equation}
\left( \sum_{i=1}^{A} \frac{1 - \alpha_{i} }{ \sigma_{\Delta t{\rm Pb},i}^2 } \right) \, t_{\rm SS} = 
\sum_{i=1}^{A} \, \alpha_{i} \, \left[ \frac{ t_{{\rm Pb},i} + \Delta t_{26,i} }{ \sigma_{\Delta t26,i}^2 }
+\frac{ t_{{\rm Pb},i} + \Delta t_{53,i} }{ \sigma_{\Delta t53,i}^2 }
+\frac{ t_{{\rm Pb},i} + \Delta t_{182,i} }{ \sigma_{\Delta t182,i}^2 } 
\right]
\end{equation}
Upon final simplication this yields
\begin{equation}
t_{\rm SS}^{*} = 
\frac{
\sum_{i=1}^{A} \alpha_{i} \left( 
 \frac{t_{{\rm Pb},i} +\Delta t_{26,i}}{\sigma_{\Delta t26,i}^2}
+\frac{t_{{\rm Pb},i} +\Delta t_{53,i}}{\sigma_{\Delta t53,i}^2}
+\frac{t_{{\rm Pb},i} +\Delta t_{182,i}}{\sigma_{\Delta t182,i}^2}
\right)
}
{
\sum_{i=1}^{A} \alpha_{i} \left(
 \frac{1}{\sigma_{\Delta t26,i}^2}
+\frac{1}{\sigma_{\Delta t53,i}^2}
+\frac{1}{\sigma_{\Delta t182,i}^2}
\right)
}.
\end{equation}


\bibliographystyle{elsarticle-harv} 
\bibliography{AnchorsAway}





\end{document}